\documentstyle[aps,eqsecnum]{revtex}
\input psfig.sty

\def\be{\begin{equation}}
\def\ee{\end{equation}}
\def\ba{\begin{eqnarray}}
\def\ea{\end{eqnarray}}

\def\sl{I\hspace*{-0.2cm}{ -}}
\def\r{{\sc r}}
\def\ve{{V_{\rm eff}}}
\begin{document}
\draft

\title{Gravity waves from relativistic binaries}

\author{Janna Levin$^{*\dagger}$, Rachel O'Reilly$^{**}$ and E. J. 
Copeland$^{**}$ }
\medskip
\address{${}^{*}$Astronomy Centre, University of Sussex,Brighton BN1
9QJ, UK}
\address{${}^{\dagger}$DAMTP, Cambridge University,
Silver St., Cambridge CB3 9EW}
\address{${}^{**}$Centre for Theoretical Physics, 
University of Sussex,Brighton BN1 9QJ, UK}

\twocolumn[\hsize\textwidth\columnwidth\hsize\csname
           @twocolumnfalse\endcsname

\maketitle
\widetext

\begin{abstract}

The stability of binary orbits can significantly shape the
gravity wave signal which future Earth-based interferometers hope to detect.
The inner most stable circular orbit has been of interest
as it marks the transition from the late inspiral to final plunge.
We consider purely relativistic orbits beyond 
the circular assumption.  
Homoclinic
orbits are of particular importance to the question of stability as they
lie on the boundary
between dynamical stability and instability.  
We identify these,
estimate their
rate of energy loss to gravity waves, and compute their gravitational 
waveforms.

\end{abstract}

\medskip
\noindent{04.30.Db,97.60.Lf,97.60.Jd,95.30.Sf,04.70.Bw}
\medskip
]

\narrowtext

Coalescing compact objects are proficient emitters of gravity
waves during the final minutes before their demise.
They are the
most popular candidate for the direct detection of gravitational radiation by 
the future LIGO/VIRGO interferometers \cite{kip}.
The gravitational wave frequencies sweep
through the LIGO/VIRGO bandwith of $1-10^4$ Hz as the inspiral gives way to
the plunge and final coalescence.
General interest in the onset of the instability to a plunge 
has focused attention on the innermost stable
circular orbit (ISCO) \cite{rdet}.
The
post-Newtonian (PN) approximation to general relativity \cite{{pn}}
has been studied in detail and 
the ISCO determined in the absence of dissipation in Ref. \cite{kww}.

There are other orbits which may be even more 
important to the question of
dynamical
stability than the ISCO. 
The homoclinic orbit in particular
arises at the boundary of the stable and unstable
manifolds in the dynamics.  
A homoclinic orbit begins and ends asymptotically close to 
an unstable
circular orbit, spiraling in and out around the center of mass.
These trajectories define the region where dynamical instability, and even
chaos, can develop \cite{bc}.
Only black holes are compact enough to capture a companion in an orbit
which approaches these
smallest radii.
We give a prescription for identifying homoclinic orbits in the 2PN expansion
in the absence of dissipation, estimate the luminosity in gravity waves,
and compute the gravitational waveform.

Long lived binaries will likely circularize by the time of inspiral
as angular momentum is lost to gravity waves.  
In dense stellar regions such as globular clusters the capture of a companion
can be an efficient mechanism for forming binaries \cite{qs}. 
There will not be sufficient time for these to circularize and we
will have to accept them and their eccentricities.  
Recently some attention has
been paid to gravity waves from more general orbits
\cite{iyer}.  We use these advances to estimate the luminosity in
gravity waves from a purely relativistic homoclinic orbit.
In order to detect the waves, very
precise theoretical templates are needed.  The raw data will then be compared
to the template over many thousands of orbits to draw out a detection.
To keep the data analysis computationally manageable, the theoretical
templates are primarily for
circular orbits.  The homoclinic orbit will chirp up in frequency as it winds
in toward the unstable circular orbit and then chirp down in frequency as it
winds back out.  To detect these, either more general templates, or new 
detection
methods, are likely required.

\section{Schwarzschild Black Holes}

Primarily for a point of comparison, we estimate the rate of
energy lost to gravitational waves and the waveforms for 
Schwarzschild black holes with a test mass companion in a relativistic
orbit.  In \S
\ref{2pnsect}
we find the energy loss and waveforms in the 2PN expansion.
Schwarzschild geodesic
motion is confined to an orbital plane and is described by the
familiar metric
	\be
ds^2=-\left (1-{2m\over \r}\right )dt^2+\left (1-{2m\over \r}\right
	)^{-1}
	d\r^2+\r^2d\phi^2
	\ee
There
are two constants of the motion; the energy $E$ and the angular
momentum $J$ per unit rest mass: 
$\left (1-{2m\over \r}\right )t^\prime =E+1 $ and 
$\r^2\phi^\prime = J$,
where a prime denotes differentiation with respect to an affine parameter.
In order to compare with the 2PN
approximation we have used the binding energy $E$ which 
is equal to the usual Schwarzschild energy minus one.
Working in Schwarzschild time $t$, the conserved angular momentum is
	\be
	\r^2\dot \phi={J\over E+1}\left (1-{2m\over \r}\right )
	\ee
and the energy
constraint equation is
	\be
	{1\over 2}\dot \r^2+V_{\rm eff}(\r)=1
	\ee
with
	\ba
	V_{\rm eff} \equiv 1&-&{1\over 2}\left (1-{2m\over \r}\right )^2
	\nonumber \\
	&+&{1\over 2(E+1)^{2}}\left (1-{2m\over \r}\right )^3\left
	(1+{J^2\over \r^2} \right )
	\ \ .
	\ea
It is well known that for a large enough angular momentum, the black hole
allows for two circular orbits; one unstable and one stable. 
The circular orbits are solutions to $V_{\rm eff}=1$.
As was done in Ref. \cite{bc}, the location, energy, and angular momentum 
of the circular orbits can be parameterized as
	\ba
	{\r_{un}/m}&=&{6\over 1+\beta} \nonumber \\
	 {\r_{st}/m}&=&{6\over
	1-\beta}\nonumber\\
	{J/m}&=&2\sqrt{3\over (1-\beta^2)}\nonumber\\
	E_{un}&=&-1+{(2-\beta)\over 3}\sqrt{2\over (1-\beta)} \nonumber\\
	m\dot \phi_{un}&=& {\sqrt{6}\over 36}(1+\beta)^{3/2}
	\nonumber 
	\ \ .
	\ea
As the angular
momentum and energy drop the two circular orbits move closer together and
eventually meet at the ISCO.  
The ISCO is strictly speaking a saddle
point of the dynamics and  
occurs for $\beta=0$ at
$\r_{un}=\r_{st}=6, m\dot \phi=\sqrt{6}/36$ 
and $J/m=2\sqrt{3},E=2\sqrt{2}/3-1$.

The ISCO is of interest as it marks the onset of a dynamical instability.
There are other orbits which may be even more 
important to the question of
the dynamical
stability. 
The homoclinic orbits are defined as those trajectories which asymptotically
approach the same unstable orbit in the infinite past as they do in the
infinite future.  These orbits lie in 
the most unstable region of phase space and
will be the site of any chaotic dynamics which may develop as a result of 
perturbations \cite{bc}.  
In the binary system they begin
infinitesimally close to an unstable circular orbit, wind out to a maximum
radius and then wind back in again just
reaching the unstable orbit.

A trajectory which begins high on an unstable circular orbit 
has enough energy to escape the pair if its energy exceeds the
asymptotic energy of a companion at rest at infinity;
that is
if $E>0$.
The homoclinic orbit can therefore only exist if
$E<0$ at the unstable circular orbit.
The
first homoclinic orbit appears when
$\beta=1/2$ and asymptotically reaches
the unstable circular orbit  at 
$\r_*/m=4,m\dot \phi_*=1/8$ and $J_*/m=4$.
Bombelli and Calzetta \cite{bc} found the homoclinic orbit explicitly
as
	\be
	{m\over \r}={(1-2\beta)\over 6}+\beta
	\tanh^2\left(\sqrt{\beta}\phi/2\right ) .\label{exact}
	\ee
with the range $0<\beta<1/2$.

\begin{figure}
\centerline{\psfig{file=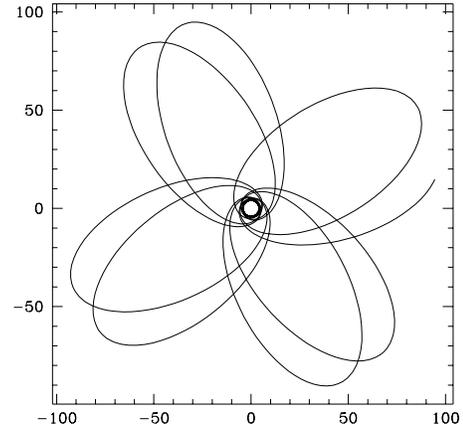,width=2.5in}}
\caption{$\beta=.47, \r=\r_{un}+0.01,\dot r=0$
\label{sch47}}  \end{figure}

No pair will ever reside exactly at the homoclinic orbit.  More
naturally, we show a nearly homoclinic orbit obtained by starting a
pair at the homoclinic $J/m$ for $\beta=0.47$ but with a
radius $0.01 m$ larger than the unstable circular orbit.
The trajectory is shown in fig.\ \ref{sch47}.  The companion executes a few
windings around the
unstable circular orbit before tracing the giant precessing ellipse.
This path is much more interesting than a simple Keplerian elliptical
orbit as is the gravitational waveform generated.

We calculate the luminosity in gravity waves and the waveform for the
homoclinic orbits in the Newtonian quadrupole approximation.
Strictly speaking, the Newtonian approximation is not appropriate for
use with gravitational sources.  In \S \ref{2pnsect}, we utilize the
waveforms properly estimated in the 2PN expansion of the two-body
problem.  The qualitative behaviour is strikingly similar to the
quadrupole approximation of this section.  The Schwarzschild solution
may also have a role in the computation of the two-body expansion.  In the
test mass limit, the 2PN expansion converges slowly to the
Schwarzschild solution.  For this reason a hybrid expansion of the
equations of motion has been proposed which is exactly Schwarzschild
in the test mass limit \cite{kww}.  A similar hybrid expansion of the
waveforms  could reduce to these Schwarzschild expressions for a
massive black hole with a light companion.

\begin{figure}
\centerline{\psfig{file=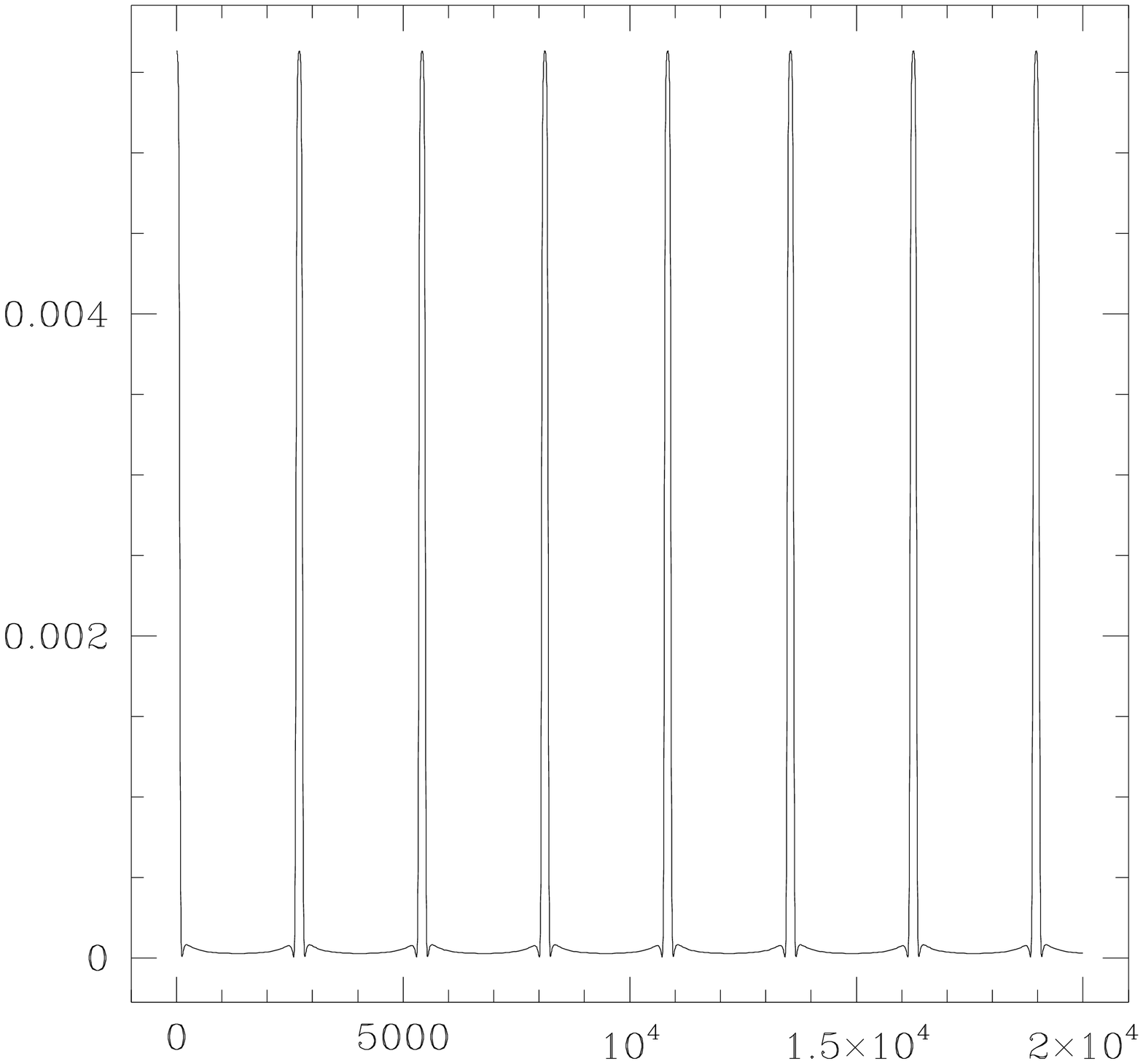,width=2.5in}}
\centerline{\psfig{file=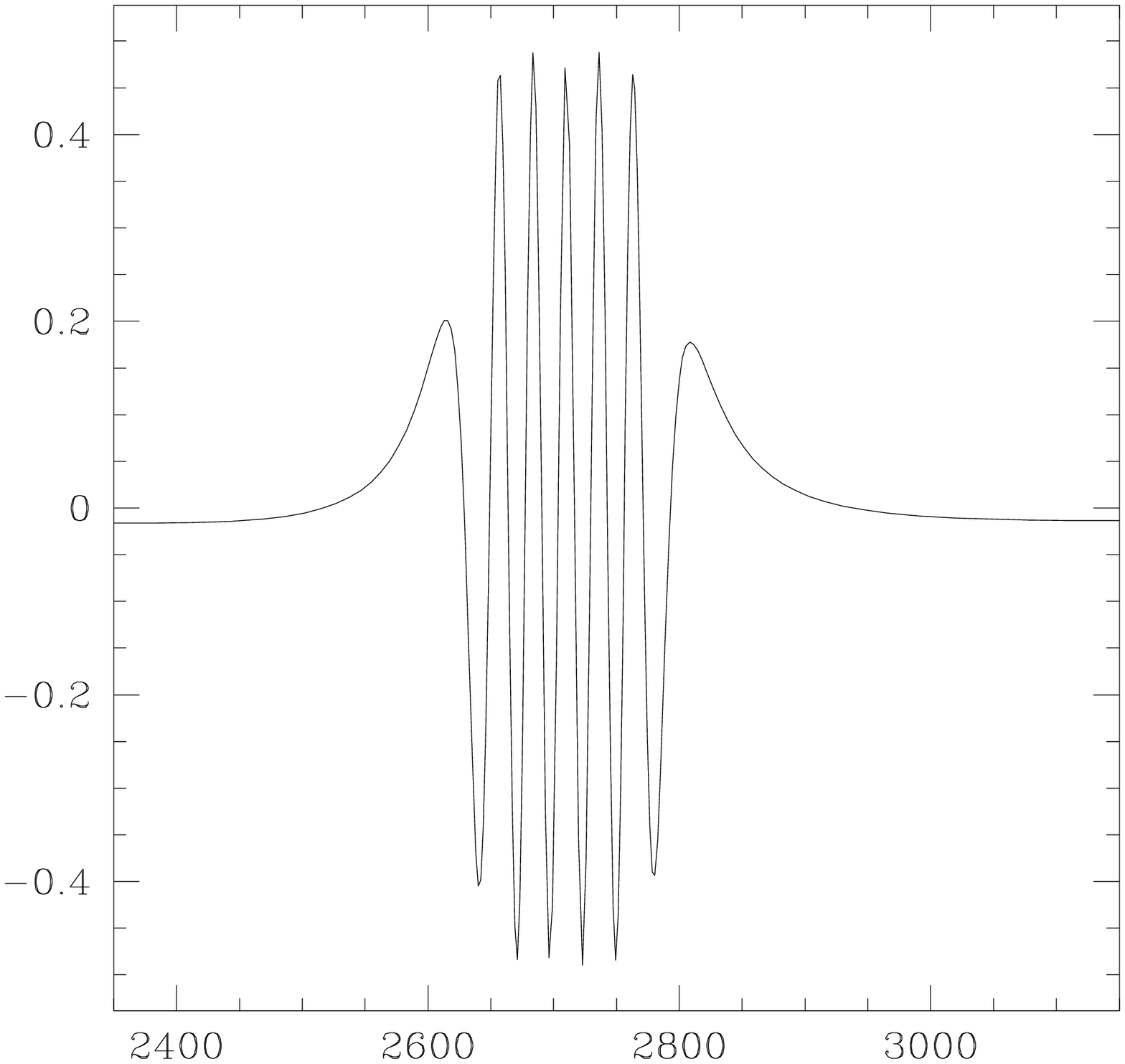,width=2.5in}}
\centerline{\psfig{file=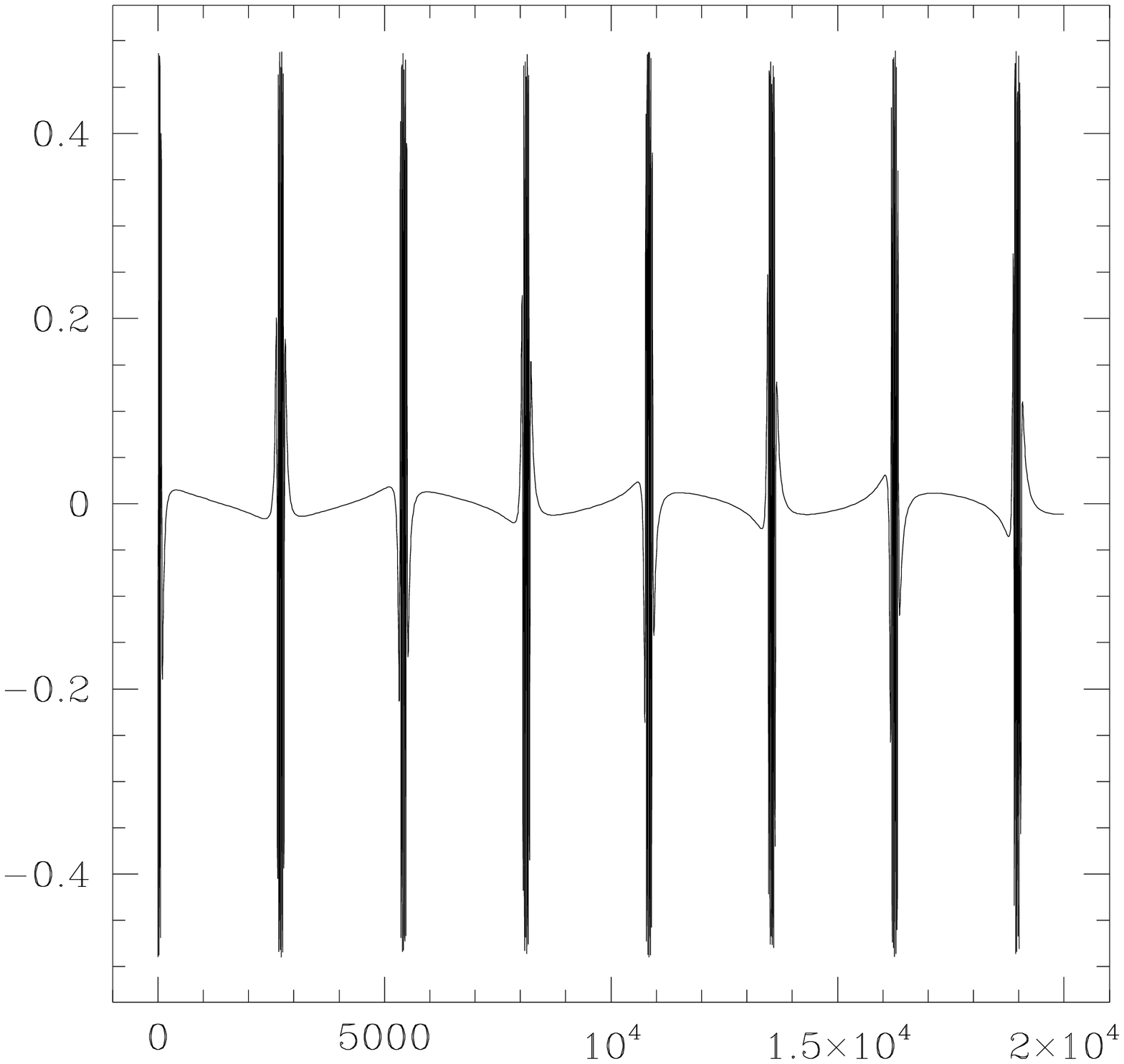,width=2.5in}}
\caption{Top:
The luminosity in gravity waves
for the orbit of fig.\ \ref{sch47} as a function of $t/m$.
Middle:  The waveform $h_{+}$ over a short interval.
Bottom: The waveform $h_{+}$ over a longer interval.
The scale of $h_+$ is arbitrary.
\label{schwv47}}  \end{figure}

The quadrupole moment is
simply $I_{i j}=mx_ix_j$:
	\ba
	I_{xx} &=& m\r^2\cos^2\phi \nonumber\\
	I_{yy} &=& m\r^2\sin^2\phi \nonumber\\
	I_{xy} &=& m\r^2\cos\phi\sin\phi\nonumber
	\ea
with all other components zero.  The luminosity in gravity waves is
estimated as
	\be
	L_{GW}={1\over 5} \left <
	\stackrel{\Large ...}{\sl}_{xx}^2+
	\stackrel{...}{ \sl}^2_{yy}+\stackrel{...}{ \sl}^2_{zz}
	+2\stackrel{...}{ \sl}^2_{xy}\right >
	\ee
with the notation
	\be
	\sl_{i j}=I_{ij}-{1\over 3}\delta_{ij}I^k_k .
	\ee
For a general orbit, this can be expressed as
	\be
	L_{GW}={1\over 5}{m^2\over 2}
	\left < Q^2+P^2+{3}W^2 \right >
	\label{lgw}
	\ee

with
	\ba
	W &=& {2\over 3}(\r\stackrel{...}{ \r}+3\dot \r\ddot \r) \\
	Q &=& 2\r\stackrel{...}{ \r}+
	6\dot \r \ddot \r-24\r\dot \r\dot \phi^2-12\r^2\dot
	\phi\ddot \phi \\
	P &=& 12 \r \ddot \r \dot \phi +12 \dot \r^2\dot \phi +12\r \dot \r
	\ddot \phi+2\r^2\stackrel{...}{ \phi}-8\r^2\dot \phi^3
	\ea
The instantaneous
rate of energy loss to gravity waves is shown in the top panel of 
fig.\ \ref{schwv47}.  As expected, the luminosity is greatest when the
binary is closest together although the pair spends a longer time at
larger separations.  For $\beta=0$, the homoclinic orbit and the
unstable circular orbit are one and the same and the luminosity
reduces to the expected circular orbit expression.
For the exact homoclinic orbit the solution eqn.\ (\ref{exact}) can be
substituted into eqn.\ (\ref{lgw}) to write the luminosity as a
function of $\phi $ only.  However, the expression is long and
involved and not particularly illuminating.  We present the numerical
results instead.

The transverse, traceless waveform can be estimated using the
Newtonian quadrupole
	\be
	h_{ij}^{TT}={2\over D}\ddot\sl^{TT}_{ij}
	\ee
with the Earth a distance $D$ from the pair in a direction $\hat N$,
$\sl^{TT}_{ij}=P_{ijkm}\sl_{km}$,
and
	\ba
	P_{ijkm}=\left (\delta_{ik}\right.&-&\left. N_iN_k\right )
	\left (\delta_{jm}-N_jN_m\right )\nonumber \\&-&{1\over 2}
	\left (\delta_{ij}-N_iN_j\right )\left
	(\delta_{km}-N_kN_m\right ).
	\ea
For the sake of illustration we place the Earth directly above the
binary
so that $\hat N=\hat z$.
The waveforms for the two polarizations are
 $h_{+}=h_{xx}=-h_{yy}$
	\ba
	h_{+}
	&=& {1\over D}\left (\ddot I_{xx}-\ddot I_{yy}\right
	)
	\\
	&=& {2m\over D}\left ( (\dot \r^2+\r\ddot \r-2\r^2\dot \phi^2)\cos 2\phi 
	-(4\r\dot \r \dot \phi+\r^2\ddot \phi)
	\sin 2\phi \right )\nonumber
	\ea

and
	\ba
	h_{\times}&=&h_{xy}=h_{yx}={2\over D}\ddot I_{xy} \\
	&=& {2m\over D}\left ( (\dot \r^2+\r\ddot \r-2\r^2\dot \phi^2)\sin 2\phi 
	+(4\r\dot \r \dot \phi+\r^2\ddot \phi)\cos 2\phi \right ).\nonumber
	\ea
The observer can be moved without complication to an arbitrary
location by using rotated quadrupole components, $\bar I=O^T I O$ with 
$O(\hat N)$ the appropriate rotation matrix.
A detector will respond to a linear superposition of $h_+$ and
$h_\times$.
The bottom panel of
fig.\ \ref{schwv47} is 
the waveform $h_+$ for the orbit of fig.\ \ref{sch47}.  The middle
panel focuses in on a shorter time interval to resolve the
oscillations as the companion sweeps near the unstable circular orbit.
This signal appears to
chirp periodically.
Although we have only used the Newtonian approximation, we will show
in the next section that the 2PN estimation of the two body problem
mimics these features with little alteration.

\begin{figure}
\centerline{\psfig{file=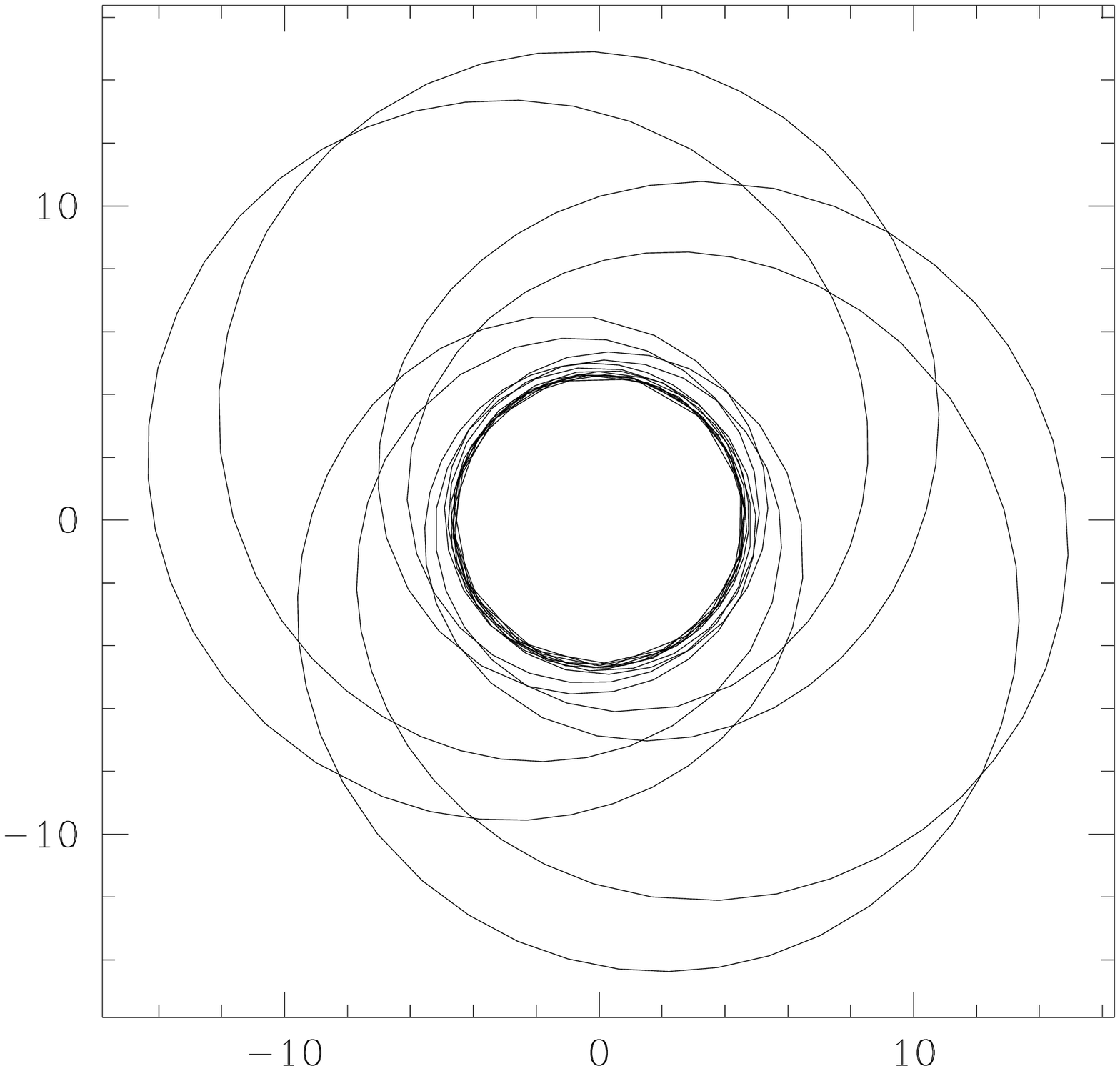,width=2.5in}}
\centerline{\psfig{file=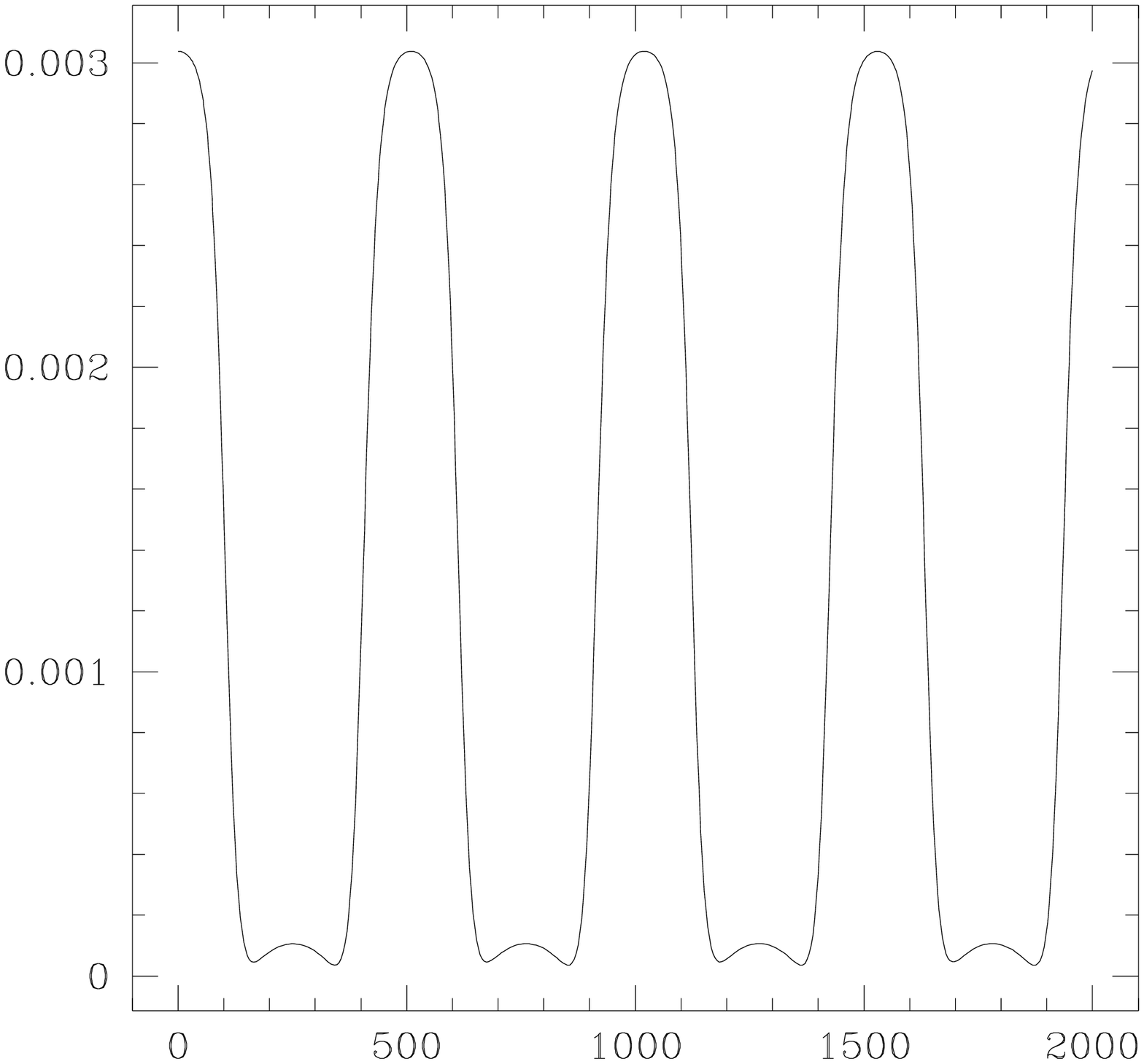,width=2.5in}}
\centerline{\psfig{file=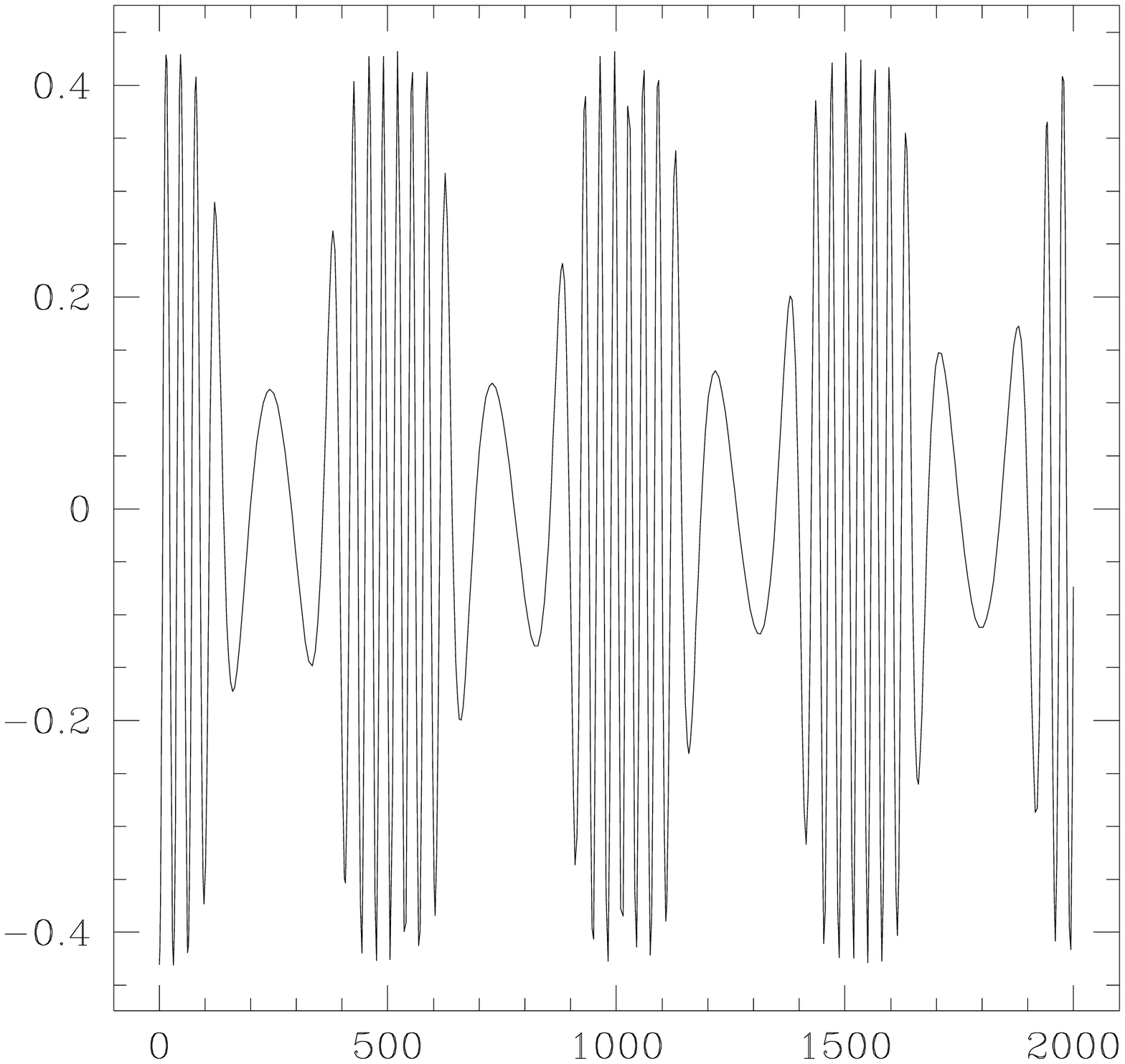,width=2.5in}}
\caption{The orbit, waveform and
the rate of energy loss for a nearly homoclinic orbit with 
$\beta=.3, \r=\r_{un}+0.01, \dot \r=0$.
\label{sch3}}  \end{figure}

We show another nearly homoclinic orbit in fig.\ \ref{sch3}
for $\beta=0.3$ which has the
angular momentum of the unstable circular orbit but begins displaced by
$0.01 m$ from the circular radius.
This orbit again winds tightly around the unstable circular orbit
before tracing a wider orbit.
The rate of energy loss to gravitational waves is shown in the middle
panel of fig.\ \ref{sch3} and the waveform $h_{+}$ is shown in the
bottom panel.

This can be compared to the waveform for an elliptical orbit
calculated in the Newtonian approximation \cite{{wal},{mp}}.
The elliptic orbits have similar peaks and drops but do not have the
multiple oscillations inbetween.

\section{The 2PN expansion}
\label{2pnsect}

\begin{figure}
\centerline{\psfig{file=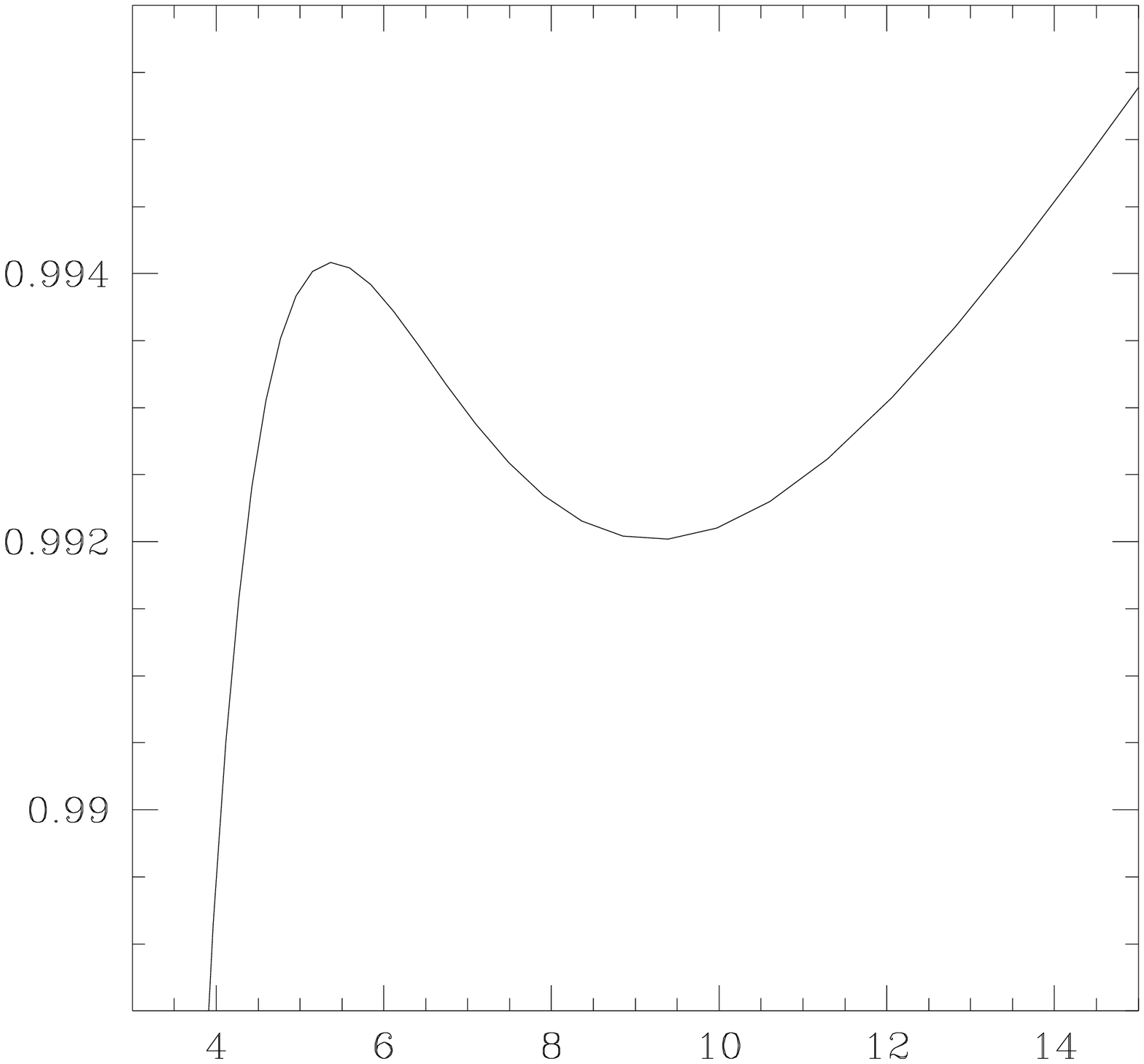,width=2.50in}}
\centerline{\psfig{file=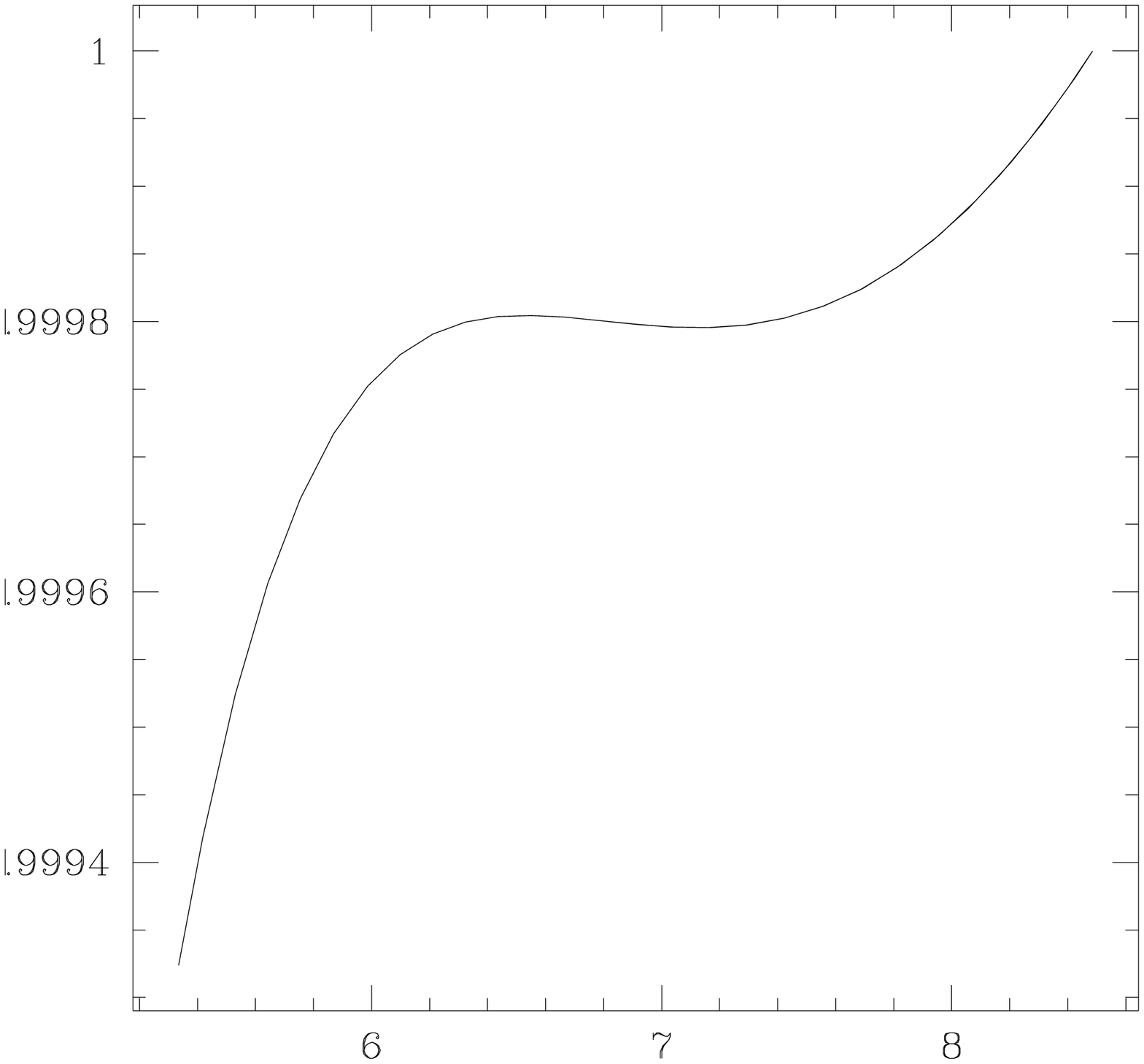,width=2.50in}}
\caption{The 2PN $\ve=1-(1/2)\dot r^2$ with $\eta=1/4$ for two
trajectories.  
Top: Initial conditions are
$r/m=20,\dot r=-0.02,m\dot \phi=0.008$.  The potential depends on both the
energy, 
$E=-0.01298$, and the angular momentum, 
$J/m=3.3067$.
Bottom: The initial conditions are
$r/m=6.8172,m\dot \phi=0.05176$ which are ISCO values for
$\eta=1/4$ but with $\dot r=0.02$ initially.
\label{typ}}  \end{figure}

The Schwarzschild solutions of the previous section can be compared
to the post-Newtonian approximations of the two-body problem.
In the 2PN expansion,
the center of mass equations of motion for the binary orbit can be
written in harmonic coordinates as \cite{{kww},{linw},{extra1}}
	\ba
	\ddot r &=&r \dot \phi^2-{m\over r^2}\left (A+B\dot
	r\right ) \label{eom1}\\
	\ddot \phi &=& -\dot \phi \left ({m\over r^2}B+2{\dot r
	\over r}\right )\label{eom2}
	\ea
where $m=m_1+m_2$ is the total mass of the pair and for 
future use we define the
product of reduced masses $\eta=m_1m_2/m^2$.  In the test mass limit, $\eta=0$,
the harmonic coordinates are related to the familiar Schwarzschild coordinates
by 
$r =\r -m$.  The form of $A(r,\dot r, \dot \phi)$ and $B(r,\dot r, \dot \phi)$
depends on the order of the PN expansion and can be found in the Appendix.
To 2PN order there are two constants of motion, the energy $E(r,\dot
r,\dot \phi)$ and the angular momentum $J(r,\dot r,\dot \phi)$ per
unit rest mass.
These elaborate expressions are also recorded in the Appendix.
The two constants of motion can be used in principle to eliminate 
$\dot r$ and $\dot \phi$ from eqn.\ (\ref{eom1}).  Just as with the 
Schwarzschild solution, the orbital dynamics can be reduced to the study of
one-dimensional motion in an effective potential,
$	\ddot r=-{\partial \ve(r)/ \partial r}$.
Integrating we have
	\be
	{1\over 2}\dot r^2 + \ve(r)={\rm constant}.
	\label{cons}
	\ee
The constant is arbitrary and we shall hereafter set it to $1$.
We will use eqn.\ (\ref{cons}) as a definition of $\ve$ in the
numerical solutions.  A typical $\ve$ for a 
black hole binary
is shown in fig.\ \ref{typ}, obtained by throwing the pair together 
at 2PN order and integrating to find 
$\ve=1-(1/2)\dot r^2$.   The shape of the potential does depend on both $E$ and
$J/m$ but nonetheless this typical $\ve$ gives an indication of the location of
two circular orbits: one unstable circular orbit near 
the maximum and one stable
circular orbit near
the minimum.

To find the homoclinic orbit, we first find all of the circular
orbits.
For circular motion,
$\ddot r_o=\dot r_o=0$.  From eqn.\ (\ref{eom1}) this demands \cite{kww}
	\be
	\quad  \dot \phi_o^2={mA_o\over r_o^3}.
	\label{cond}
	\ee
We solve the circular orbit condition (\ref{cond})
for $\dot \phi_o(r_o)$ as a function of the
radius of the circular orbit and the ratio of masses $\eta$.
The solution is provided in the Appendix eqn.\ (\ref{phi2})
for reference and provides the complete initial conditions for
circular orbits.
We substitute $\dot\phi_o$ into $J$ to define an angular momentum
along circular orbits
	\be	
	\bullet \quad  J_o(r_o)\equiv J(r_o,\dot r_o=0,\dot \phi_o)
	\ee
and into $E$ to define an energy along circular orbits
	\be	
	\bullet \quad  E_o(r_o)\equiv E(r_o,\dot r_o=0,\dot \phi_o)
	.
	\ee
These are
functions of $r_o$ only,
once $\eta$ is
specified. 
We find the smallest circular radius, $r_*$, for which
	\be
	\bullet \quad   E_o(r_*)<0.
	\ee
The first homoclinic orbit begins at this unstable circular orbit.
As the angular momentum and the energy
decrease,
the last homoclinic orbit coincides with the ISCO.
All homoclinic orbits can therefore be found by 
finding $r_*$ and $r_{\rm ISCO}$
as functions of $\eta$.

\begin{figure}
\centerline{\psfig{file=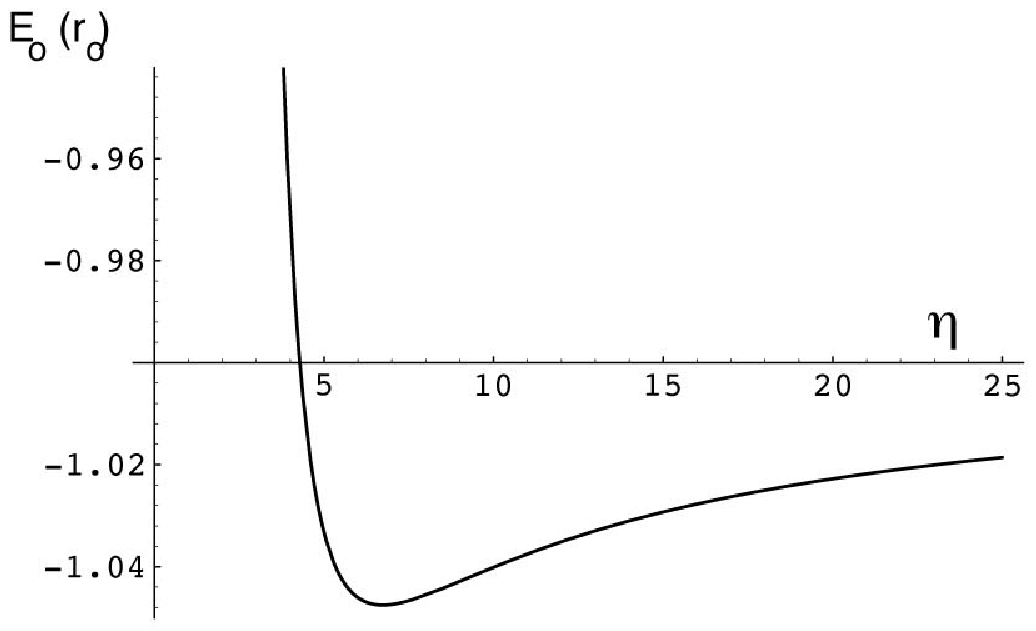,angle=0,width=3.in}}
\centerline{\psfig{file=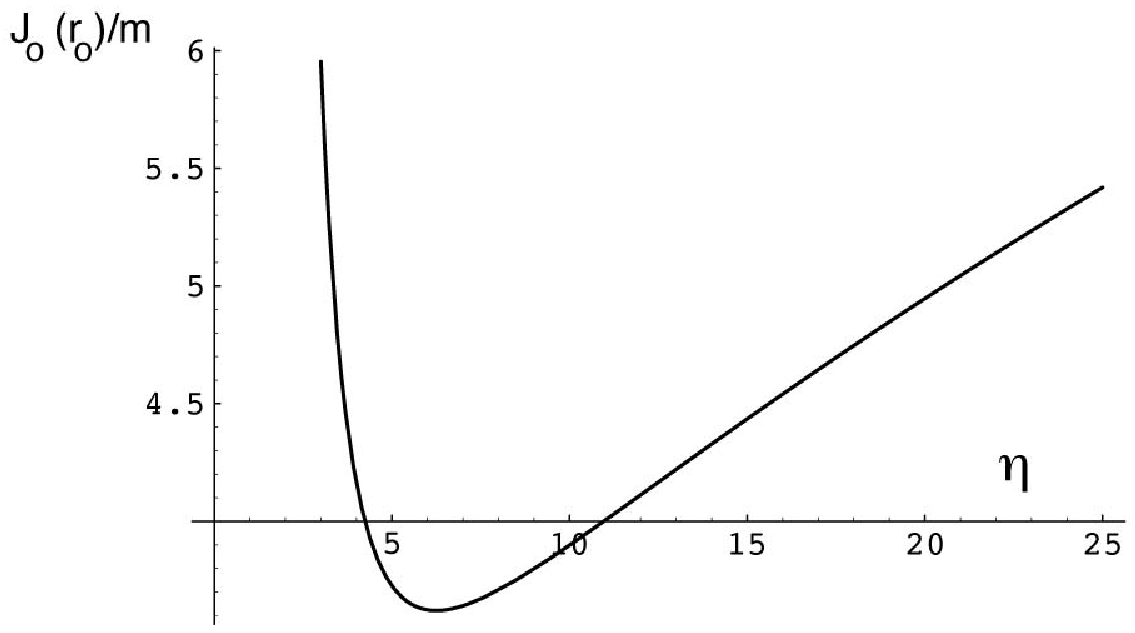,angle=0,width=3.in}}
\caption{The energy along circular orbits (top) and the
angular momentum along circular orbits (bottom)
in the 2PN expansion with $\eta=0.1/(1.1)^2$.
\label{jbig}}  \end{figure}

\begin{figure}
\centerline{\psfig{file=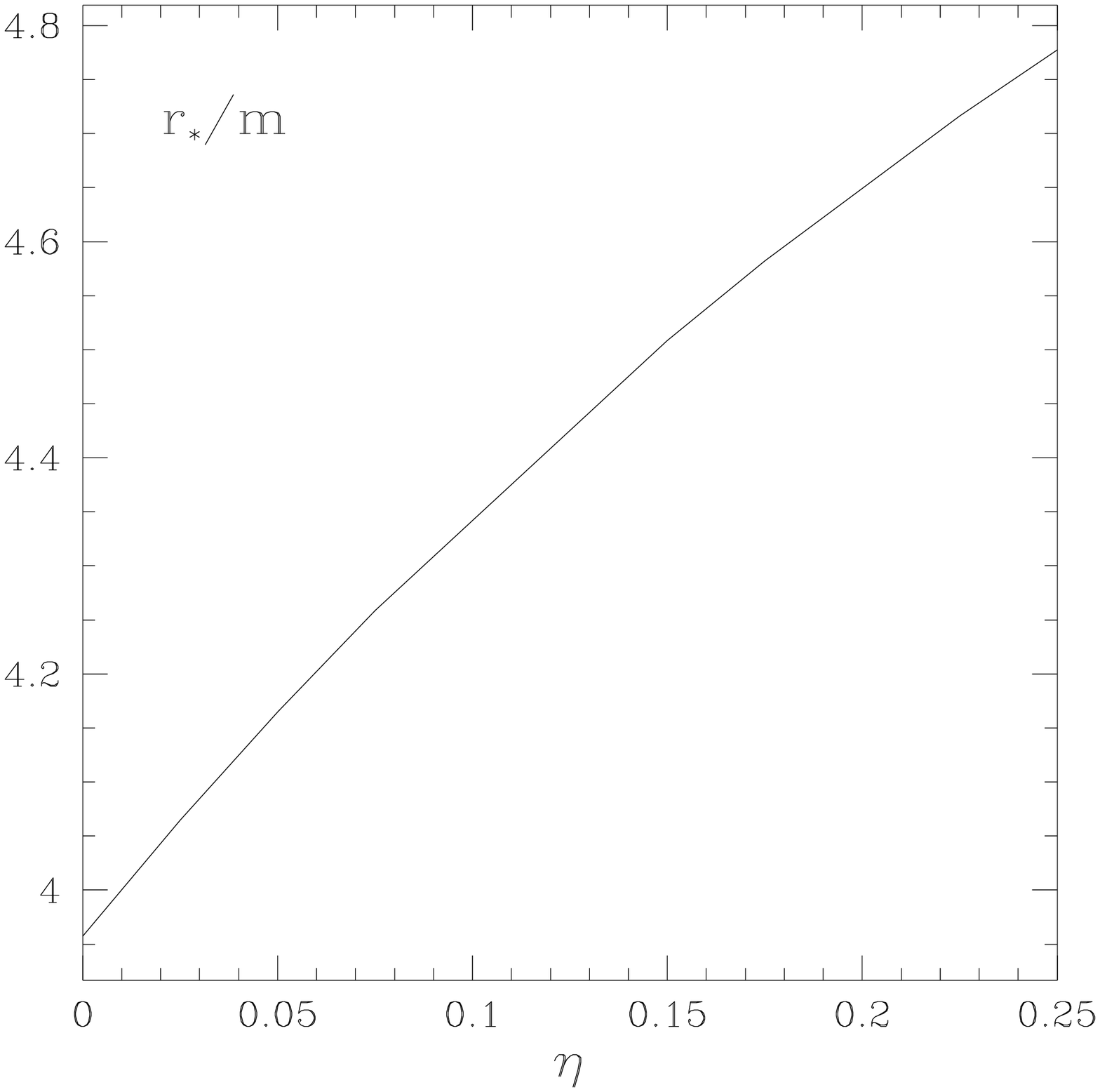,width=2.25in}}
\centerline{\psfig{file=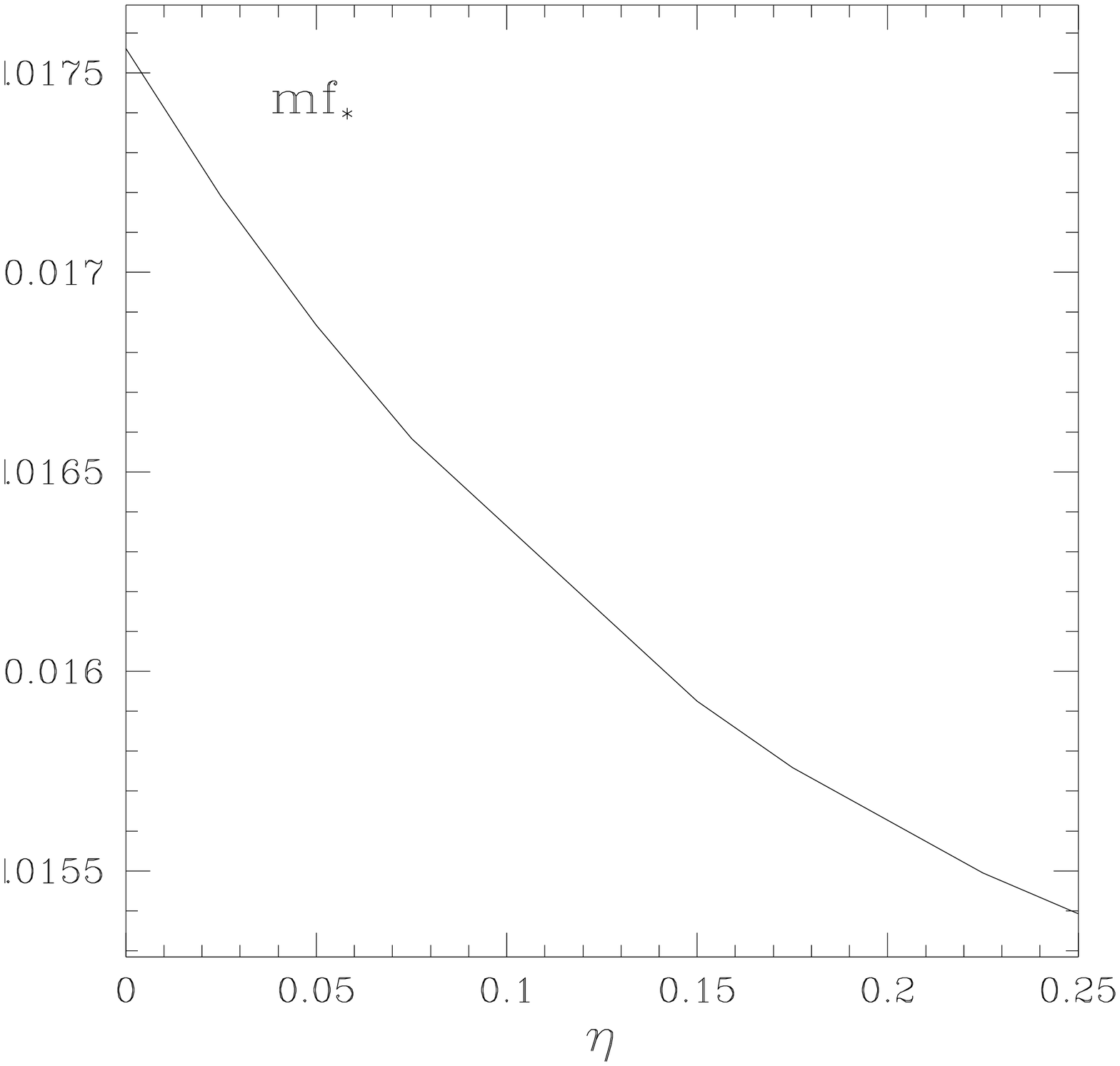,width=2.25in}}
\centerline{\psfig{file=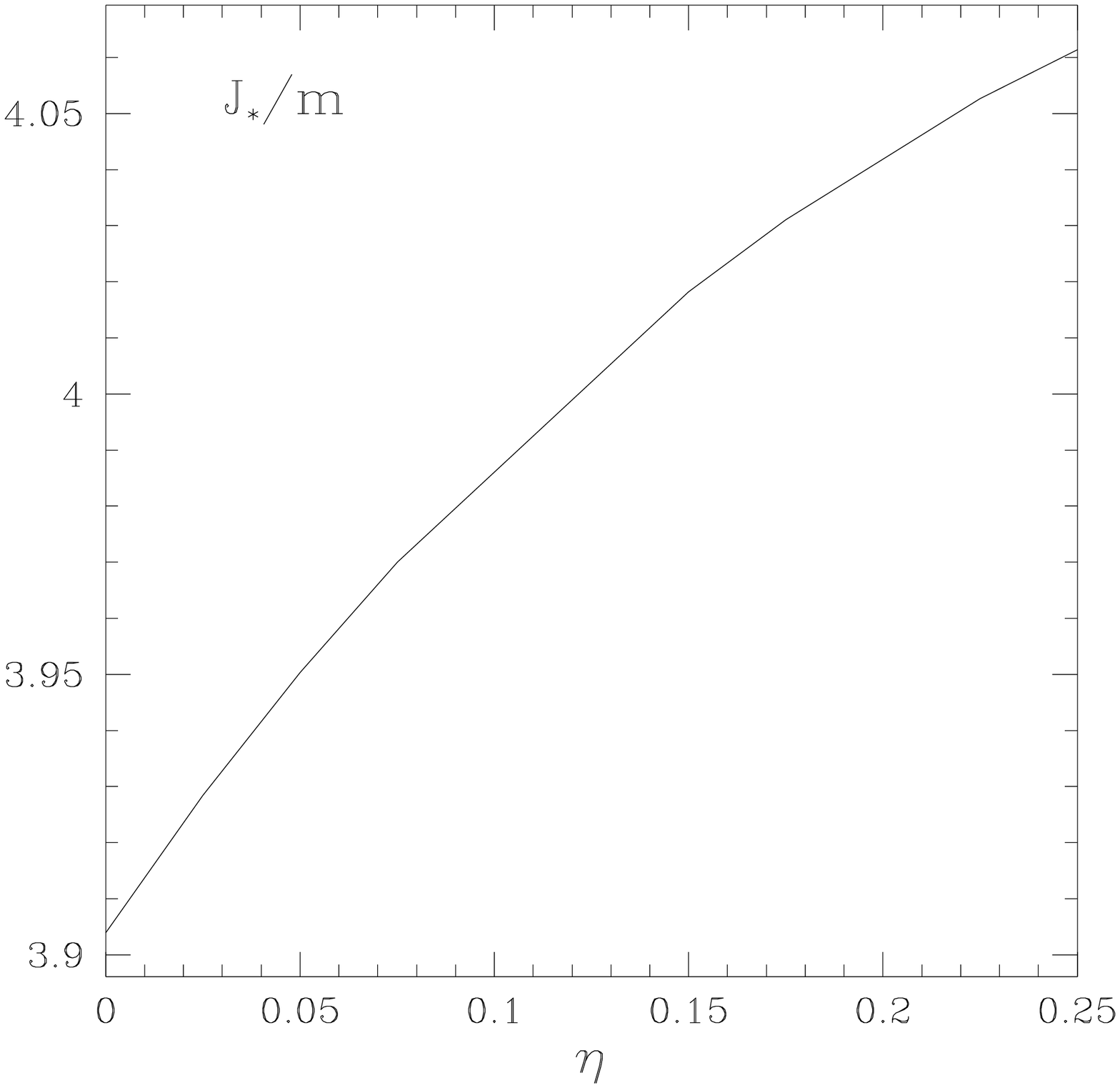,width=2.25in}}
\caption{The radius (top), the orbital 
frequency 
(middle), and angular momentum (bottom)  of the smallest
unstable circular orbit for which there is a homoclinic orbit as
functions of $\eta$ in the
2PN expansion.  The energy by definition is zero.
\label{star}}  \end{figure}

For a companion of mass $1/10$th its partner, 
$\eta=0.1/(1.1)^2$ then $E_o(r_o)$ and
$J_o(r_o)/m$ are shown in fig.\ \ref{jbig}.
For unstable circular orbits greater than $r_{*}/m=4.2866$,
$E<0$.  
This can be repeated for any $\eta$ to generate figs.\
\ref{star}.
These show the values of $r_*/m$, the orbital frequency $f_*=\dot \phi/(2\pi)$,
and $J_*/m$ for 
the unstable circular orbit when $E_*=0$ as functions of $\eta$.

We follow Ref.\ \cite{kww} to locate the ISCO.
The stability of the fixed points is tested by perturbing eqns.\
(\ref{eom1})-(\ref{eom2})
about $(r_o,\dot r_o=0,\dot \phi_o)$ to obtain,
	\be
	{d\over dt}\pmatrix{\delta r\cr  \delta \dot r\cr
	 \delta \dot \phi}
	=\pmatrix{0 & 1 & 0 \cr
	a & 0 & b \cr
	0 & c & 0}
	\pmatrix{\delta r\cr \delta \dot r\cr
	\delta \dot \phi}
	\label{mat}
	\ee
where
	\ba
	a &=&3\dot \phi_o^2-{m\over r_o^2}
	\left ({\partial A\over \partial r}\right )_o  \nonumber \\
	b&=&2r_o\dot \phi_o -{m\over r_o^2}\left (	
	\partial A\over \partial \dot \phi\right )_o
	\nonumber \\
	c&=&-\dot \phi_o
	\left ({2\over r_o} +{m\over r_o^2}
	\left (\partial B\over \partial \dot r\right )_o\right )
	.
	\label{abc}
	\ea
The perturbations obey 
	\be
	\pmatrix{\delta r\cr \delta \dot r\cr
	\delta \dot \phi}\propto e^{i\lambda t}
	\ee
with $i\lambda $ the eigenvalues of (\ref{mat}),
	\be
	\lambda=0,\quad \lambda=\pm(-a-bc)^{1/2}
	.
	\ee
Stable oscillations about a circular orbit correspond to real
$\lambda$ so that $a+bc<0$.  The ISCO is the minimum value of 
$r_o$ that has a real $\lambda $ and satisfies condition (\ref{cond})
\cite{kww}.

\begin{figure}
\centerline{\psfig{file=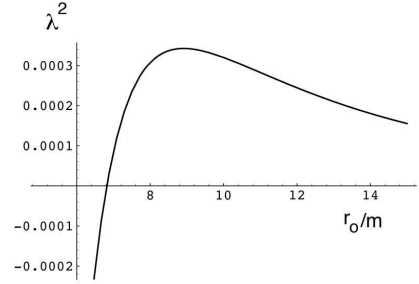,angle=0,width=3.in}}
\caption{$\lambda^2$ along circular orbits 
in the 2PN expansion with $\eta=1/4$.
\label{lam}}  \end{figure} 

\begin{figure}
\centerline{\psfig{file=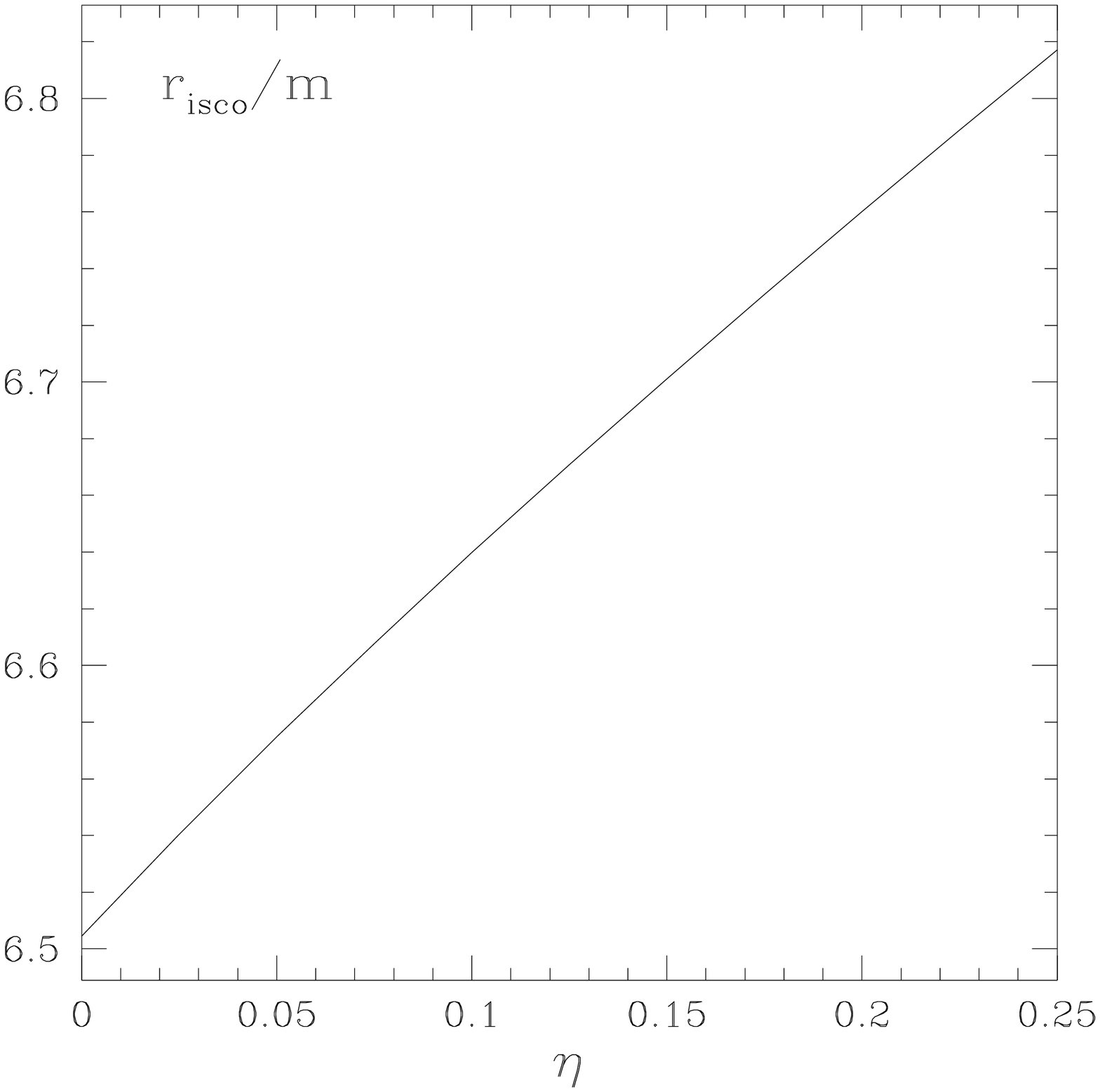,width=2.25in}}
\centerline{\psfig{file=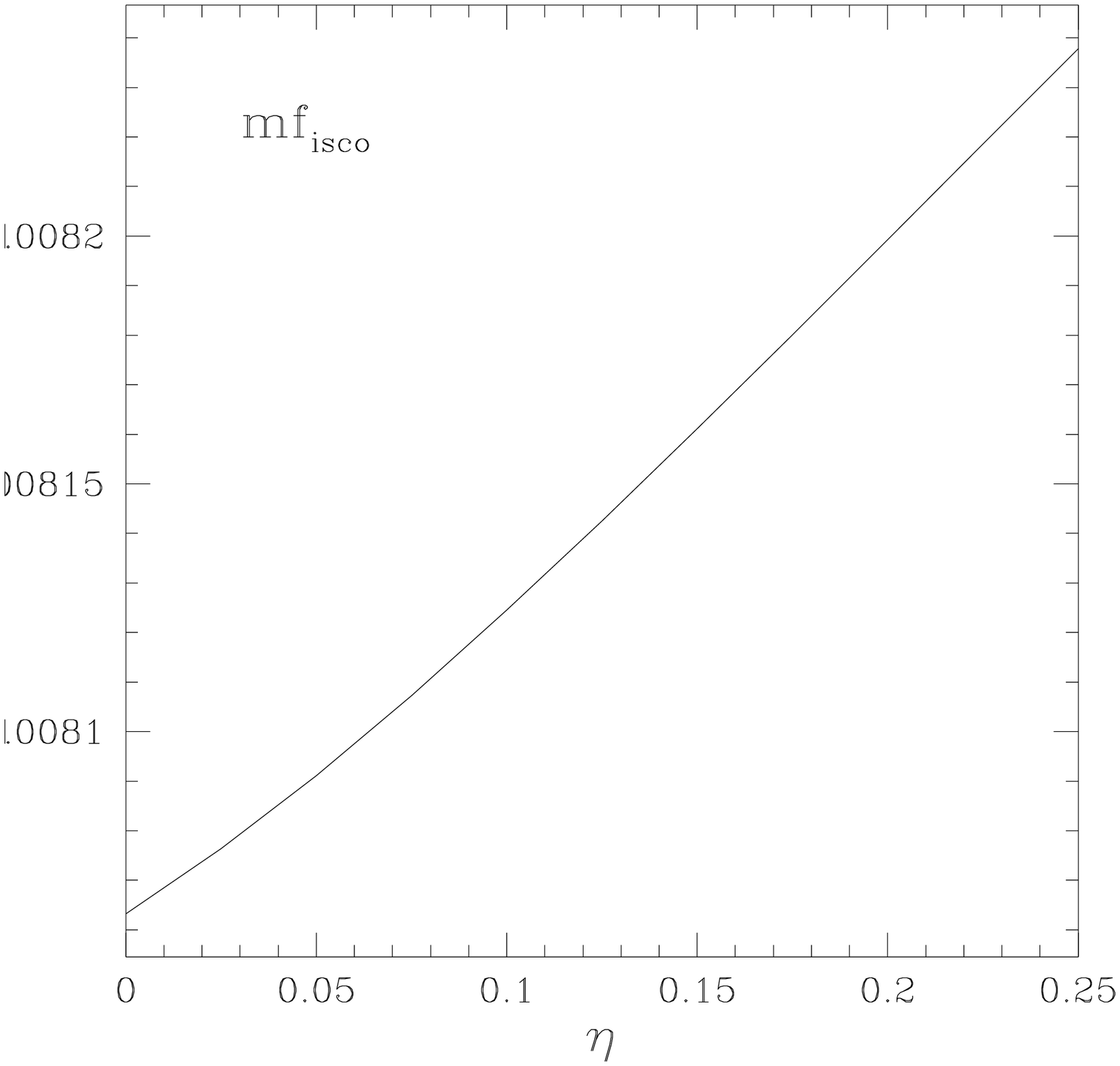,width=2.25in}}
\caption{The radius (top) and frequency (bottom) of the ISCO as
functions of $\eta$ in the
2PN expansion.
\label{isco}}  \end{figure}

\begin{figure}
\centerline{\psfig{file=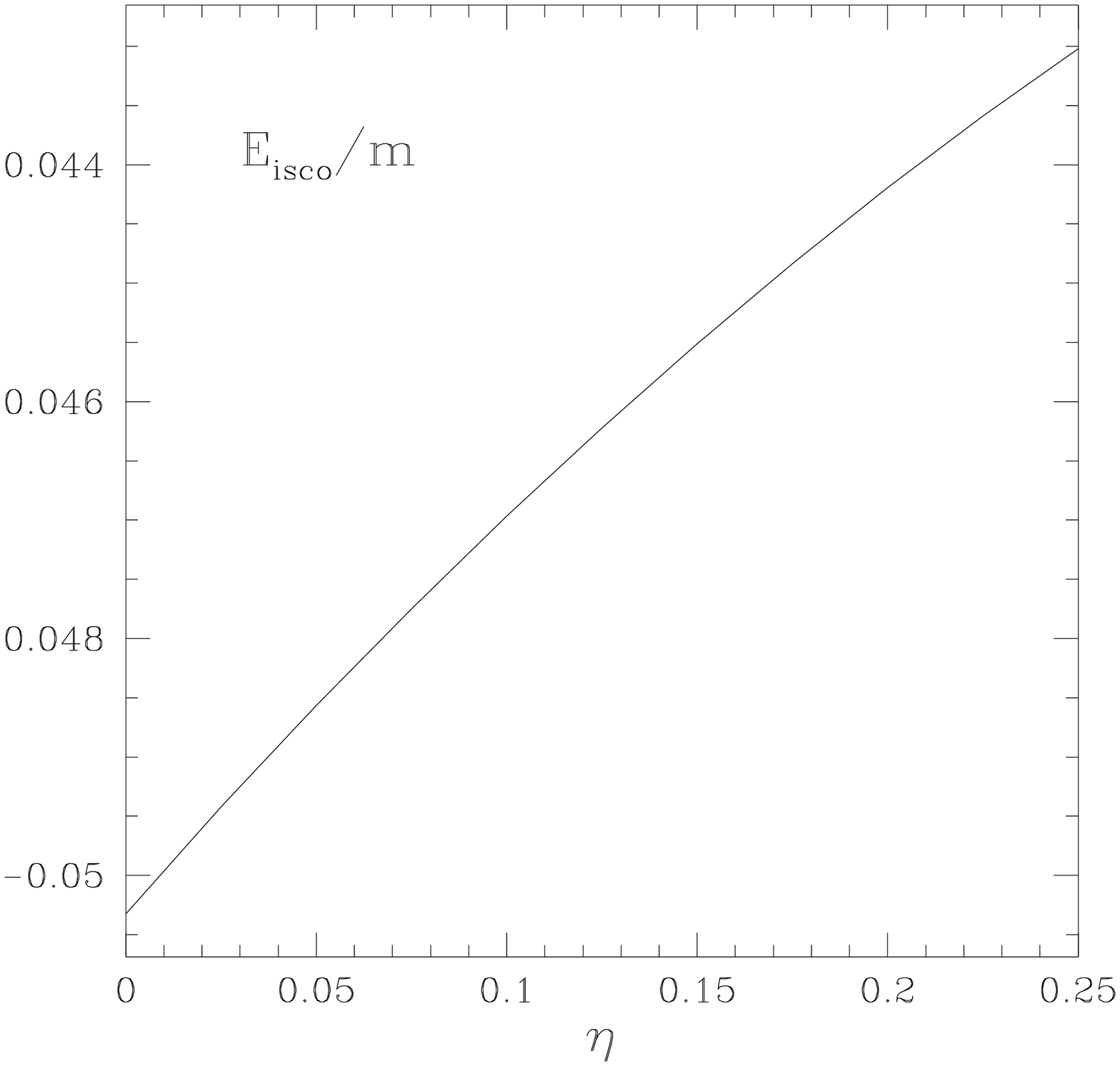,width=2.25in}}
\centerline{\psfig{file=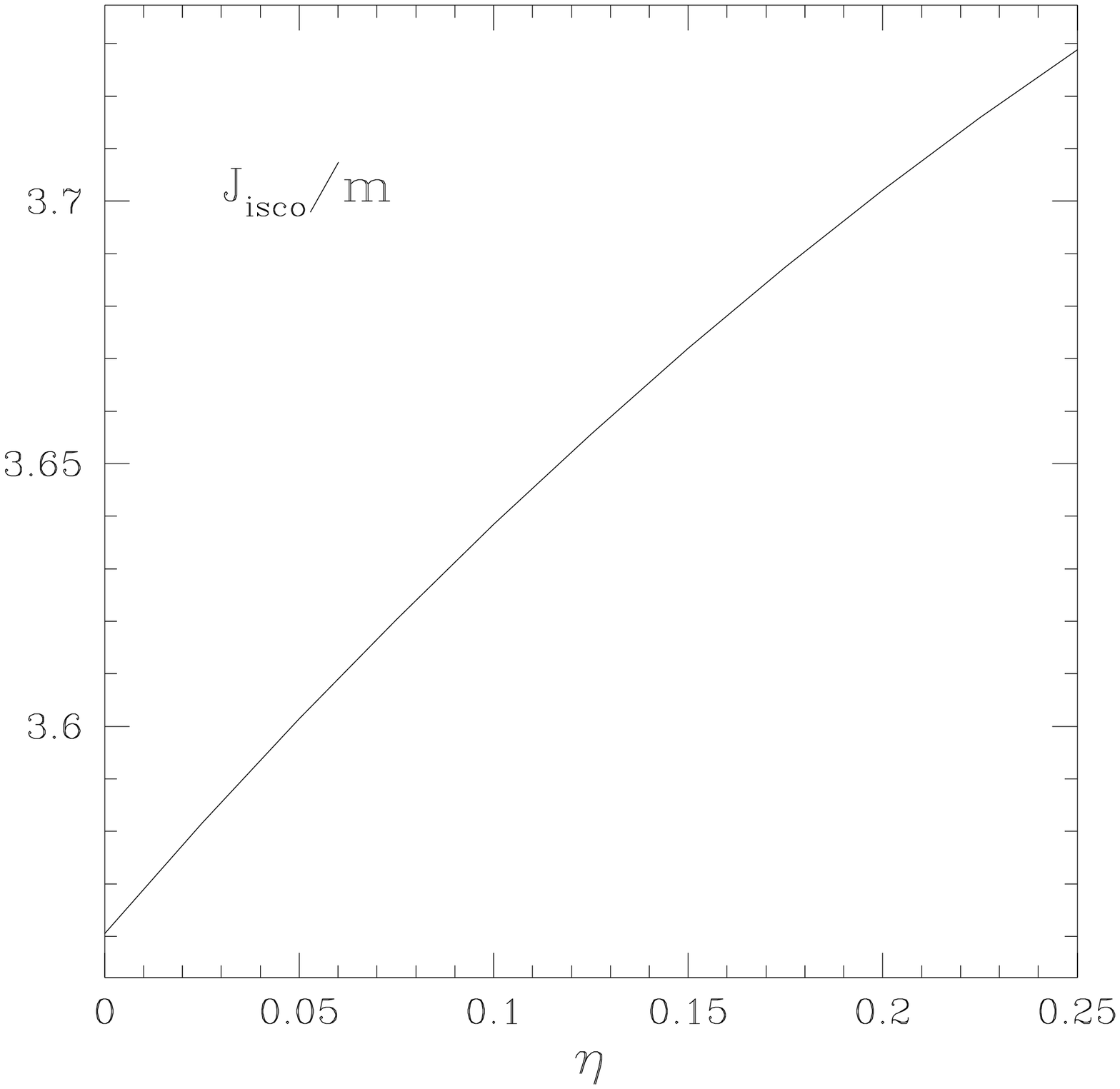,width=2.25in}}
\caption{The energy (top) and the angular momentum 
(bottom) of the ISCO as functions of $\eta $ in the 2PN expansion.
\label{je}}  \end{figure}

Using eqn.\ (\ref{a}) for $A$ and 
eqn.\ (\ref{b}) for $B$, we find 
$\dot \phi_o(r_o)$ in eqn.\ (\ref{phi2}).
The resultant $\lambda^2(r_o)$ 
is illustrated in fig. (\ref{lam}) for an equal mass companion.
For any $\eta$, the values of $r_{\rm isco}$, the orbital frequency
$f_{\rm isco}=\dot \phi/(2\pi),E_{\rm isco}$ and
$J_{\rm isco}$ are shown in figs. \ref{isco}-\ref{je}.

With the ISCO and the critical unstable circular orbit we have the
initial conditions for any homoclinic orbit in the 2PN expansion.

\subsection{Nearly homoclinic paths}

\begin{figure}
\centerline{\psfig{file=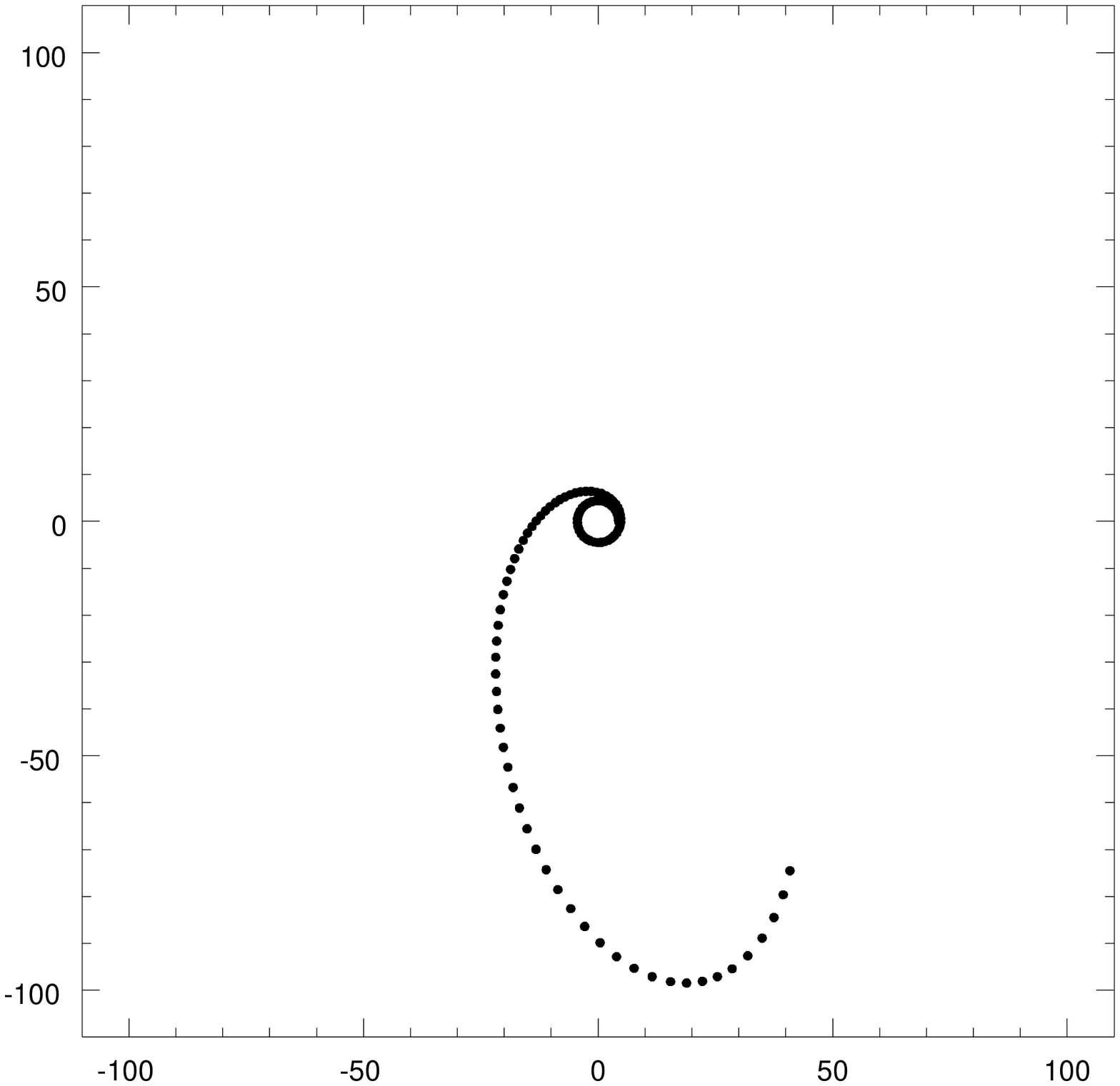,width=2.2in}}
\centerline{\psfig{file=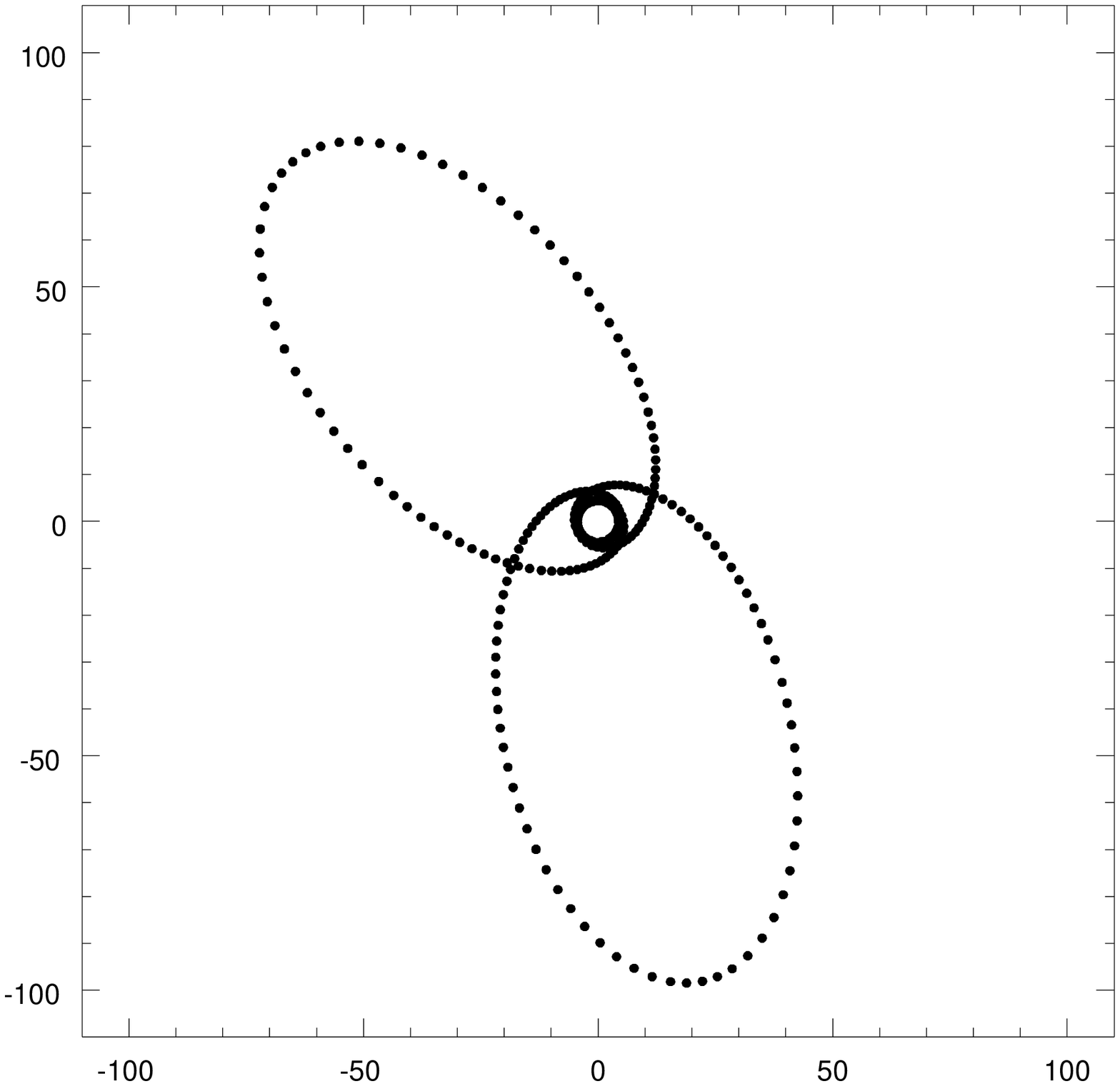,width=2.2in}}
\centerline{\psfig{file=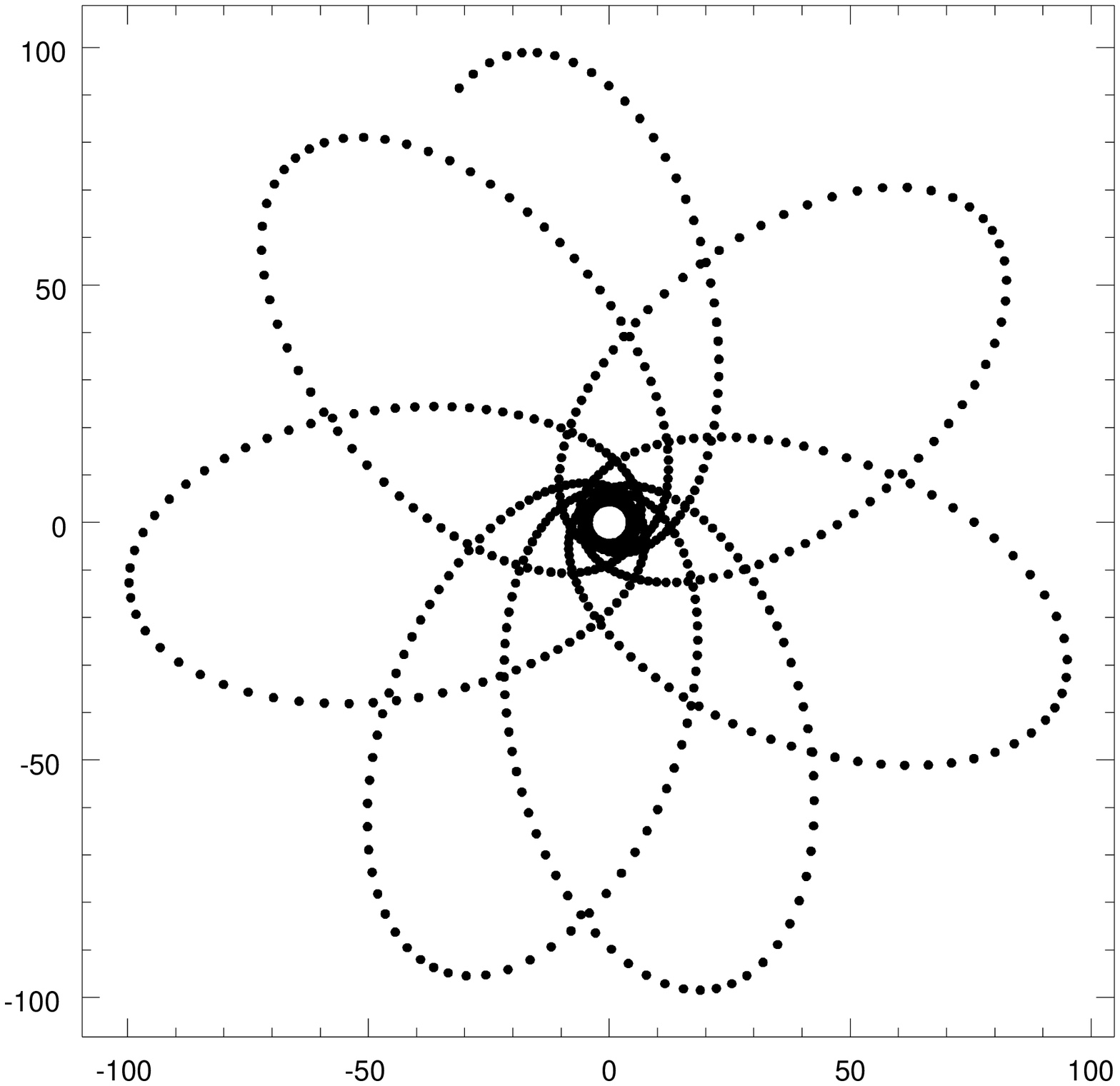,width=2.2in}}
\caption{A series of snap shots
for a nearly homoclinic orbit.
The ratio of masses is $m_2/m_1=0.1$ so that $\eta=0.082644628$.
The initial conditions were
$r/m=4.409$, $\dot r=0$, $m\dot \phi=0.0985406$.  Consequently
$E=-0.0084324135$ and $J/m=3.9166041778$.
\label{orb1}}  \end{figure}

\begin{figure}
\centerline{\psfig{file=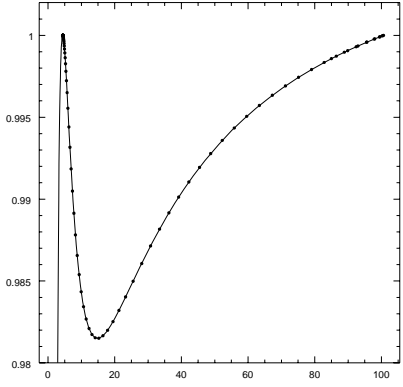,width=2.25in}}
\caption{The solid line is the full potential while the dots indicate
that part of the potential traced out by the relativistic orbit.
\label{v1}}  \end{figure}

\begin{figure}
\centerline{\psfig{file=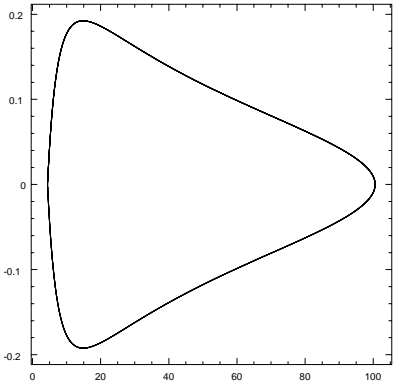,width=2.25in}}
\caption{The phase space $(r,\dot r)$ for the trajectory of fig.\ \ref{orb1}.
\label{rdr1}}  \end{figure}

An example of a nearly homoclinic orbit 
for $\eta=0.1/(1.1)^2$ is obtained numerically by
starting on an unstable circular orbit at $r/m=4.409$ and 
knocking the orbit 
slightly outwards.
The binary orbit is not just a simple precessing
ellipse.  The orbit first glides around the unstable circular orbit,
then executes a giant ellipse, swings
around the
inner unstable circular orbit again before tracing another
ellipse in a nearly opposite direction.  This overall interleafed
tracing then precesses.  This orbit is relativistic and is not even 
{\it quasi}-Keplerian.
Snapshots of the orbit are shown in fig.\ \ref{orb1} and fig.\
\ref{v1} shows the effective potential.
A phase space diagram shows periodicity in $(r,\dot r)$ in fig.\
\ref{rdr1}.

\begin{figure}
\centerline{\psfig{file=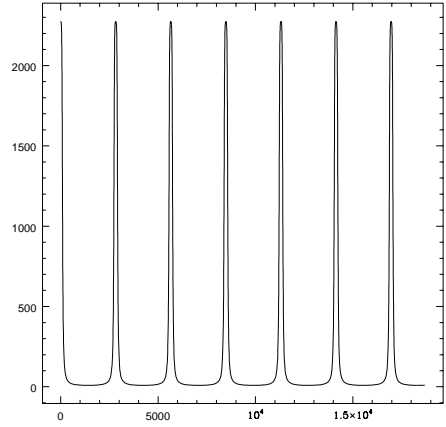,width=2.5in}}
\caption{The orbital frequency of eqn.\ (\ref{fgeq})
measured in Hz as a function of $t/m$.
\label{fg1}}  \end{figure}

The gravitational wave frequency measured by a distant observer is 
roughly twice
the orbital frequency for a circular orbit.
For elliptical orbits the waves oscillate at once, twice and three
times
the orbital frequency \cite{mp}.
Since the orbital frequency here is not constant, the comparison is
more ambiguous.  
The waveform oscillates
roughly twice per winding around the center of mass.
The orbital frequency in units of $Hz$ is
	\be
	\left (m\over 1.4 M_\odot\right )
	f_{}={\dot \phi\over 2\pi}\left ({c^3\over 1.4  G M_\odot}
	\right )\ {\rm Hz} ,
	\label{fgeq}
	\ee
and is shown in fig.\ \ref{fg1}.  
For a pair with a
total mass twice $1.4M_\odot$, the frequency $\sim {\cal O}(2000)$
Hz at the circular orbit and drops to
${\cal O}(10)$ Hz.  Although
these frequencies are still within the LIGO range,
the signal will appear to chirp up and then down again.  
For black hole pairs with larger total mass, the
signal will move in and out of LIGO's bandwith.

\begin{figure}
\centerline{\psfig{file=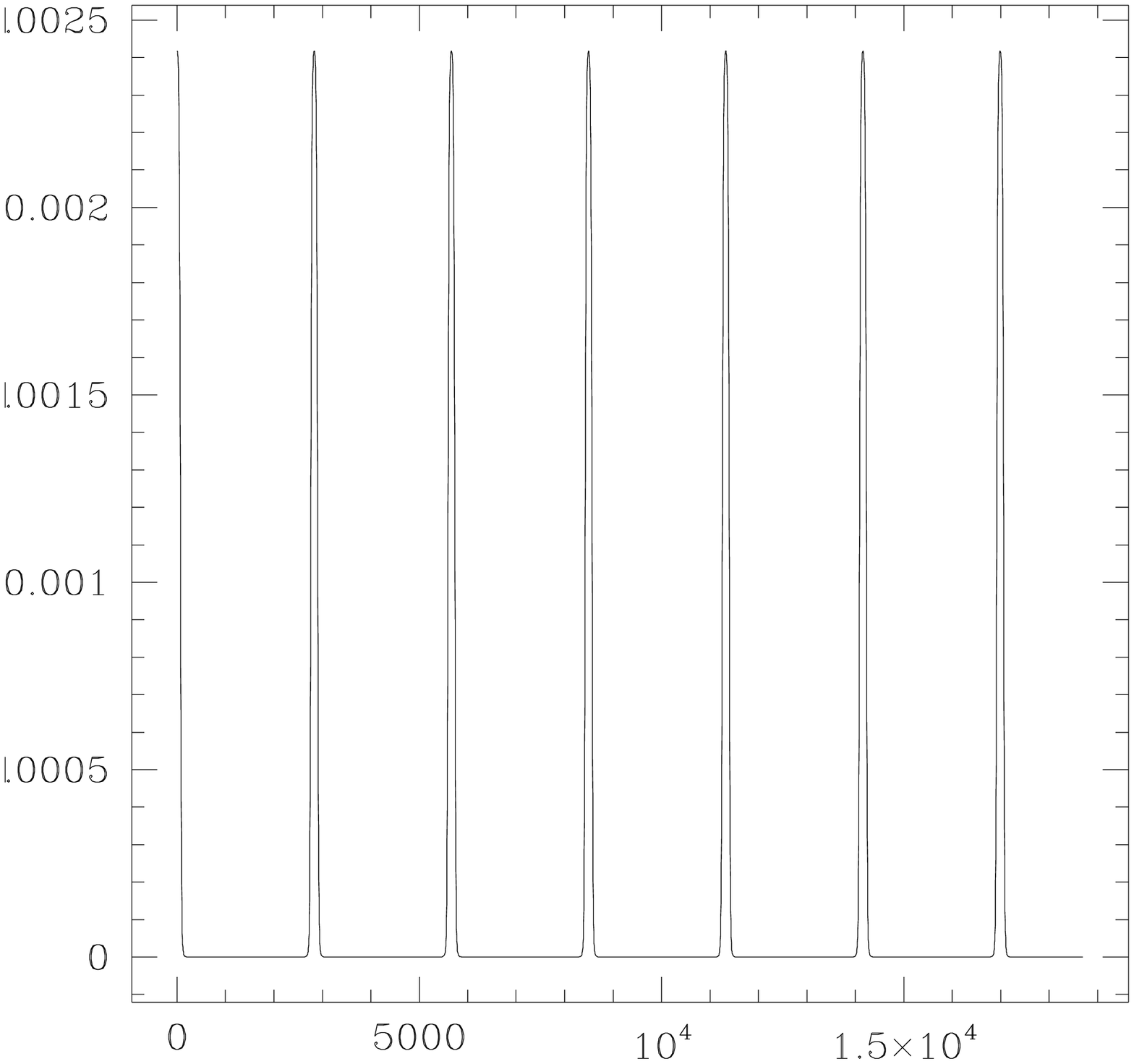,width=2.5in}}
\centerline{\psfig{file=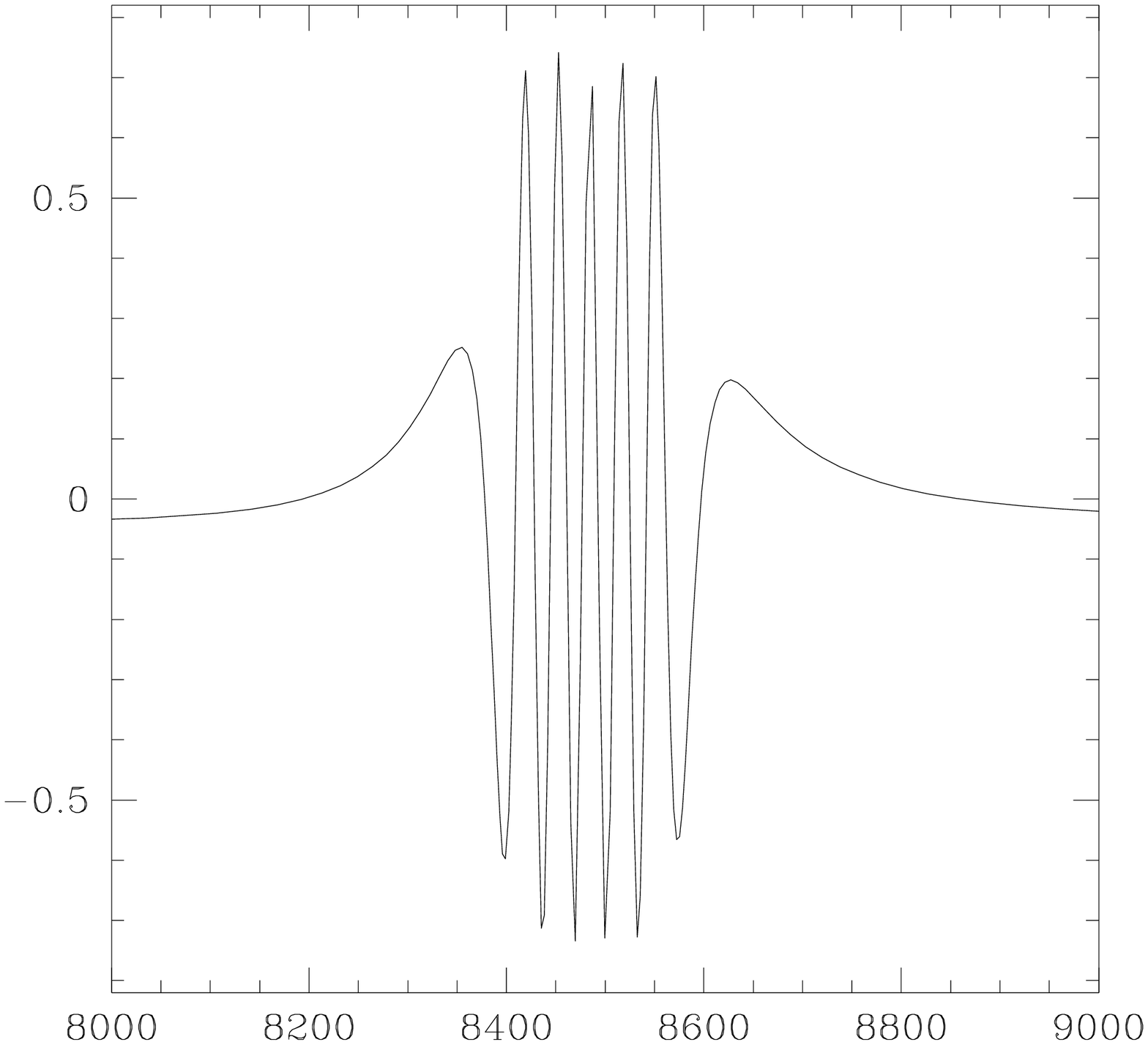,width=2.5in}}
\centerline{\psfig{file=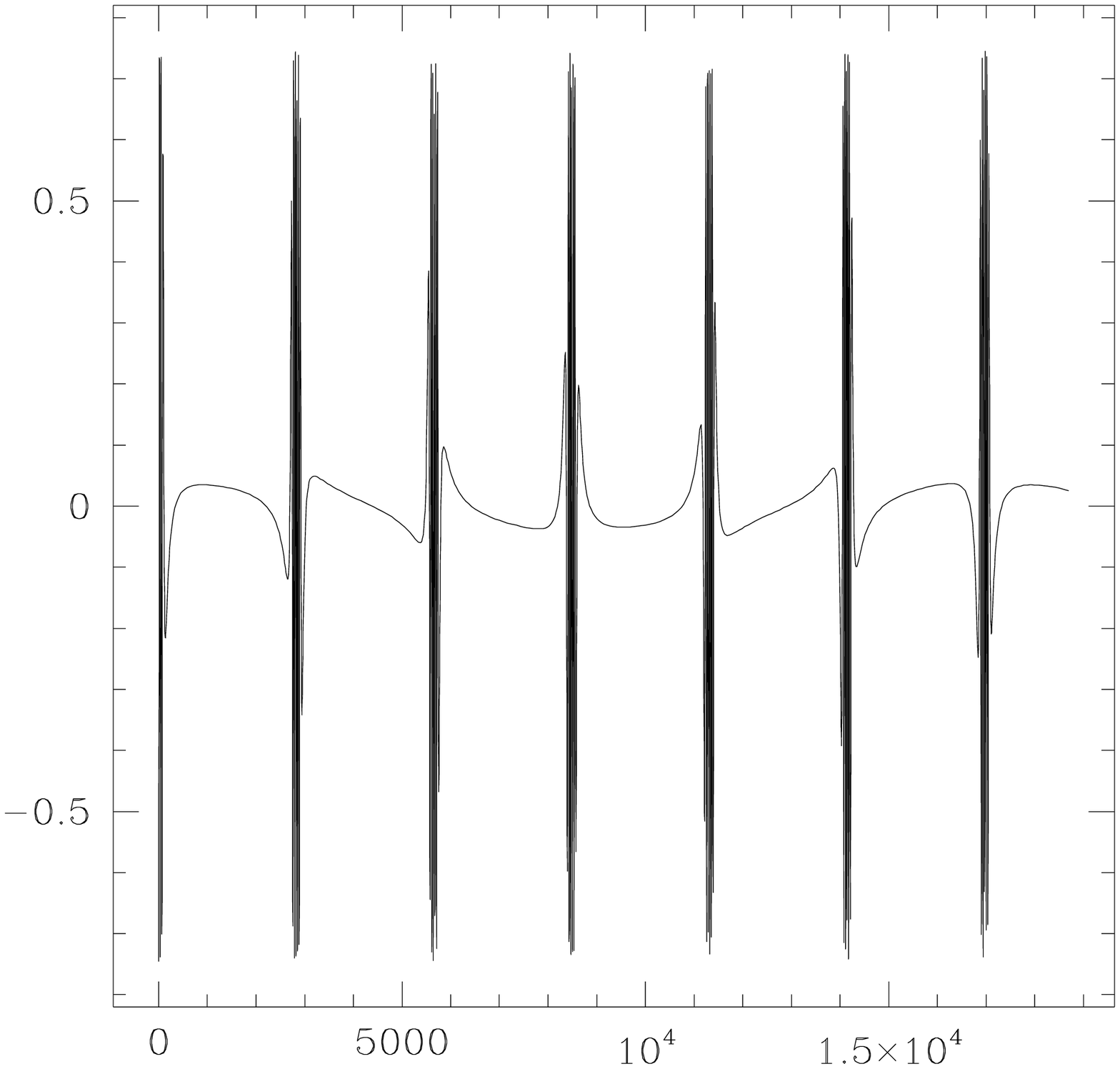,width=2.5in}}
\caption{Top:
The rate of energy loss to gravitational waves
for the orbit of fig.\ \ref{orb1} as a function of $t/m$.
Middle:  Resolution of the waveform $h_{+}$ for a short interval.
Top:  The waveform $h_{+}$ over the entire interval of
the orbit of fig.\ \ref{orb1}.  The scale is arbitrary.
\label{wv1}}  \end{figure}

It is unclear if a circular orbit template integrated against data which chirps
in and out will lock onto a signal.
The far-zone energy and angular momentum flux is calculated in Ref.\
\cite{iyer} for general binary orbits in the 2PN formalism.  We use
their results (quoted in the Appendix) to draw 
$L_{GW}$
in the top panel of fig.\ \ref{wv1}.  As expected the rate of energy 
loss is greatest near the unstable circular orbit but the
orbit spends much more time emitting gravity waves at the lower rate.

Also shown is the waveform along this orbit using the 2PN
approximation to the far-zone waveform of Ref.\ \cite{iyer}.
We have written the results of Ref.\ \cite{iyer} for $h^{TT}_{km}$ in
the Appendix eqn.\ (\ref{htt}).
Notice how strikingly similar the 2PN waveform is to the Newtonian
approximation for the waveform from the similar
Schwarzschild orbit of figs.\ \ref{sch47} \& \ref{schwv47}.

\begin{figure}
\centerline{\psfig{file=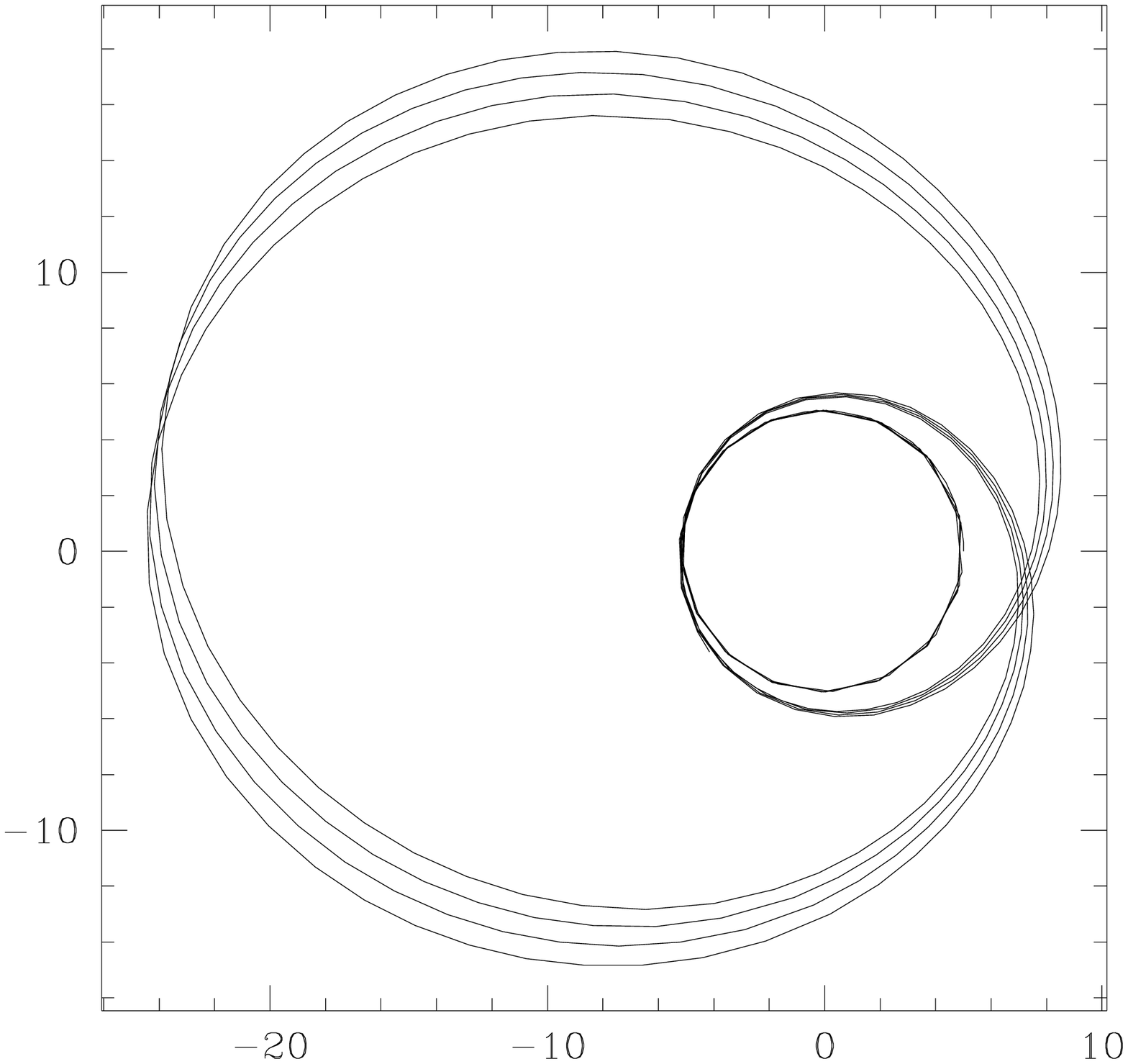,width=2.5in}}
\centerline{\psfig{file=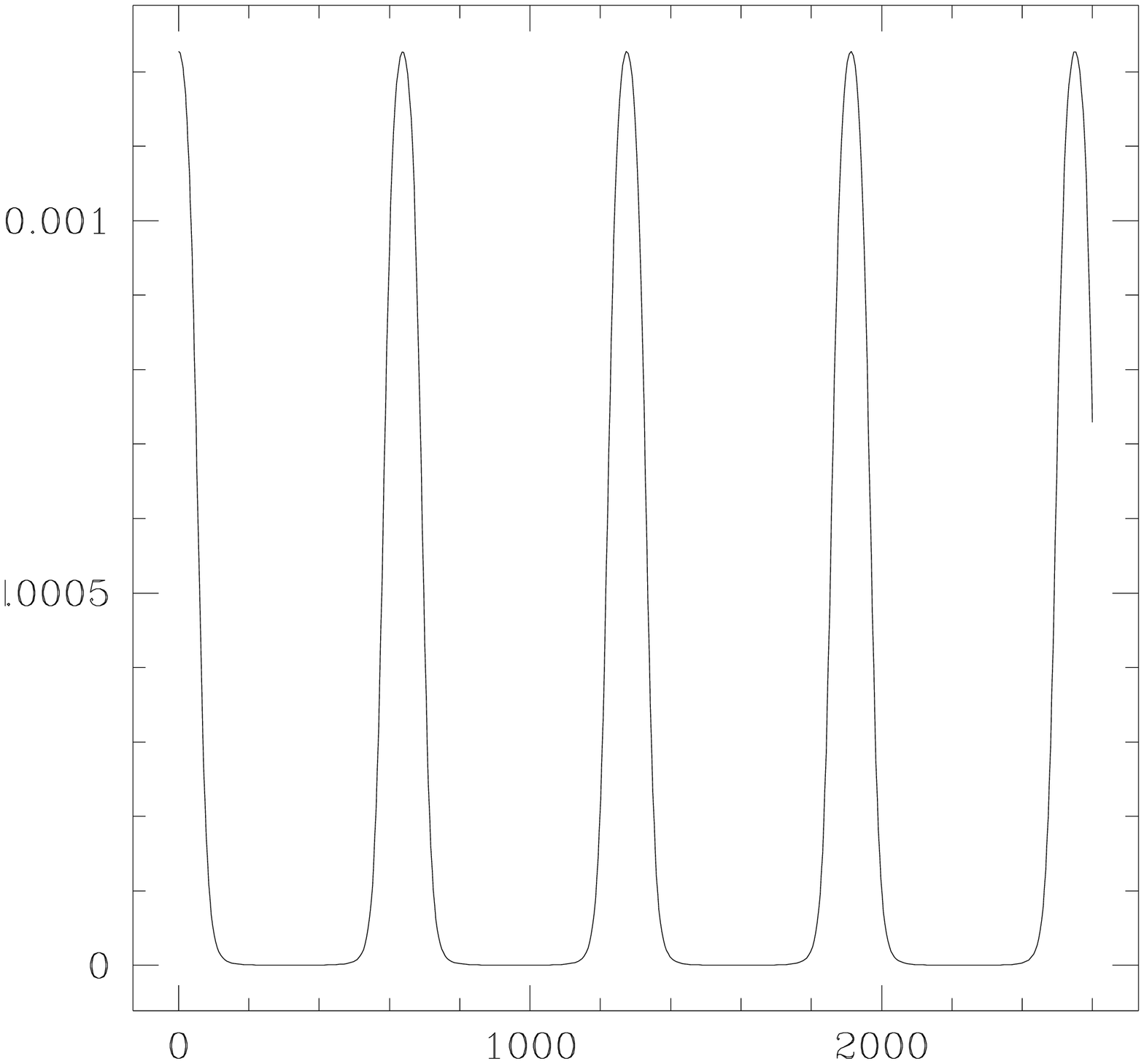,width=2.5in}}
\centerline{\psfig{file=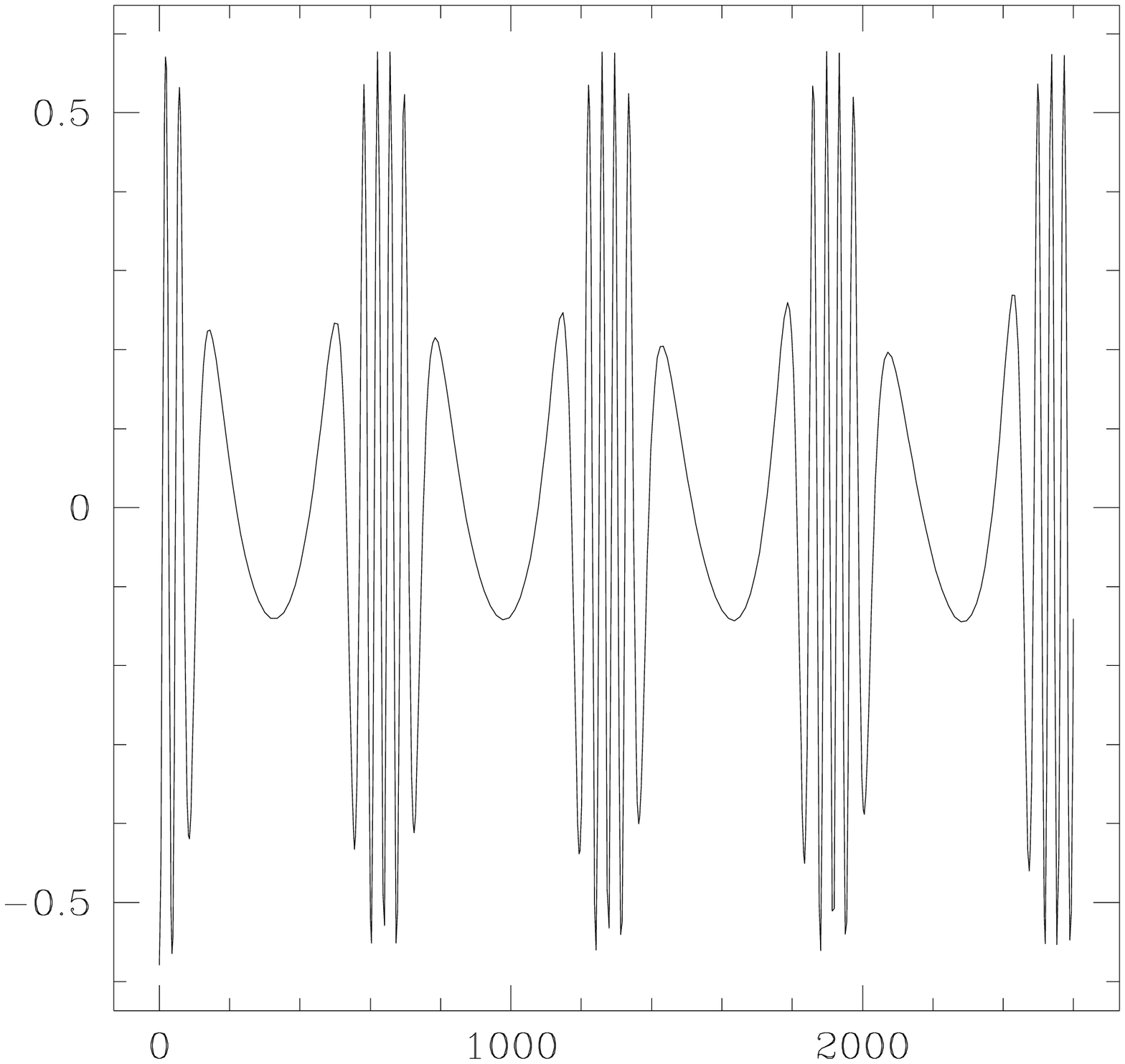,width=2.5in}}
\caption{Middle:  Top:  Another nearly homoclinic orbit with 
initial values as in eqn.\ (\ref{ic3}).
The rate of energy loss to gravitational waves
for the above orbit.  Bottom:  The waveform $h_{+}$.
\label{wv3}}  \end{figure}

We find another nearly homoclinic orbit for equal mass binaries so
that
$\eta=1/4$.
We begin the orbit at
	\be
	r_i/m = 5.01	\quad \quad
	m\dot \phi_i = 0.0886 .
	\label{ic3}
	\ee
These initial conditions are the same as 
the corresponding unstable circular orbit but with initial radius
$0.01 m$ greater.
The orbit, luminosity in gravity waves, and the waveform $h_+$ are
shown in fig.\ \ref{wv3}.
Notice that the waveform $h_+$ oscillates several times 
during the sweep around the unstable circular orbit.
Again, there is a striking similarity between the 2PN waveforms and
the waveforms calculated in the quadrupole approximation for the
Schwarzschild orbit of fig.\ \ref{sch3}.

It is difficult to say how generic these relativistic orbits are in a
realistic capture scenario.  Clearly eccentric orbits are more likely
than circular ones.  For angular momenta in the range
$J_{\rm isco}<J<J_o$, these relativistic orbits are most natural.  If
a black hole captures a companion in an initially eccentric
quasi-Keplerian orbit.  The angular momentum and energy lost to
gravity waves will not be sufficient to circularize prior to plunge
but may drive the pair into the range $E< 0$ and $J<J_o$ and therefore
onto a winding relativistic orbit.

Although the frequencies are within the LIGO detection bandwith,
these waveforms may not be detectible by the method of matched
filtering against circular templates.  Eccentric signals matched
against circular filters result in a loss in signal-to-noise
\cite{mp}.  The chirping up and down of these relativistic signal makes
them even more difficult to pick up with a circular template.  A
quantitative analysis of the loss in signal-to-noise is currently underway.

The effect of the
radiative back reaction on the instability 
is also important.
Dissipation due to radiative back reaction appears
at 5/2PN-order.
The location of the circular orbits is uneffected by these terms.
The circular orbits still obey eqn.\ (\ref{cond})
and $A(r_o,\dot r_o=0,\dot \phi_o)$ is the same with or without 
radiative back reaction
since the 5/2PN contribution to $A$ (eqn.\ (\ref{52}))
is proportional to $\dot r$.
The stability of a given $r_o$ will however be affected.  Dissipation
will likely make orbits more unstable and intuition would suggest
the ISCO would move outward.
A binary on a nearly homoclinic path 
may plunge together as a result
of energy loss which pushes the pair beyond the unstable asymptote.
Alternatively, they
may fall into a decaying oscillation about the stable point, gradually
spiraling together.
Which outcome results will depend on the initial conditions of 
the specific orbit.

Even more intriguing is the possibility that
the gravity waves disrupt the orbit to the point of chaos.
A hypothetical gravitational wave
can, under certain conditions,
lead to 
a stochastic region along the homoclinic orbit \cite{bc}.
The importance  of this is clear.  
Chaos may be an 
inevitable element in both the coalescence and the gravity wave
spectrum
\cite{me}.
However, the strength, frequency, and waveform
specific to the binary may not have the right form to induce chaos.
With the waveforms for general orbits 
of Ref.\ \cite{iyer} and the homoclinic orbits we have
isolated, 
a more realistic search for chaos in coalescing binaries
is possible.

\bigskip
\section*{Acknowledgments}

We thank Bernard Schutz for insightful comments in the early stages of this
work.  
We gratefully acknowledge PPARC support.  EJC thanks the Isaac Newton
Institute for hospitality.

\vfill\eject

\appendix

{\hsize\textwidth\columnwidth\hsize\csname
           @twocolumnfalse\endcsname
\widetext
\section{2PN Equations of Motion, Luminosity and Waveforms}

The center of mass equations of motion for the 
relativistic binary can be written 
as
	\ba
	\ddot r &=&r \dot \phi^2-{m\over r^2}\left (A+B\dot
	r\right ) \nonumber \\
	\ddot \phi &=& -\dot \phi \left ({m\over r^2}B+2{\dot r
	\over r}\right )\nonumber 
	\ea
\cite{{kww},{linw},{extra1}}.
The equations for $A,B,E$ and $J$ can be found in Ref.\ 
\cite{{kww},{linw},{extra1}} and 
are included here for the purposes of keeping the paper self-contained.
In the PN expansion, the expression for $A$ and $B$ can be obtained
order by order in the small quantity $v^2=\dot r^2+r^2\dot\phi^2$.
To 2PN order 
	\be
	A=1+A_1+A_2
	\label{a}
	\ee
with

\begin{eqnarray}
A_1 & = & -2(2+\eta)\frac{m}{r}+(1+3\eta)v^2-\frac{3}{2}\eta\dot{r}^2 
	\nonumber \\
A_2 & = &
	\frac{3}{4}(12+29\eta)\left(\frac{m}{r}\right)^2+\eta(3-4\eta)v^4+\frac{15}{8}\eta(1-3\eta)\dot{r}^4
	\nonumber \\
&-&
	\frac{3}{2}\eta(3-4\eta)v^2\dot{r}^2-\frac{1}{2}\eta(13-4\eta)\frac{m}{r}v^2
	\nonumber	\\
	&-& (2+25\eta+2\eta^2)\frac{m}{r}\dot{r}^2\nonumber
\end{eqnarray}
and 
	\be 
	B=B_1+B_2
	\label{b}
	\ee
with 
\begin{eqnarray}
B_1 & = & -2(2-\eta)\dot{r} \nonumber \\
B_2 & = & -\frac{1}{2}\dot{r} \left
[
\eta(15+4\eta)v^2-(4+41\eta+8\eta^2)\frac{m}{r}-3\eta(3+2\eta)\dot{r}^2
\right] \nonumber
	\end{eqnarray}
The subscripts refer to the order of the term in the PN expansion.
To 2PN order, there is no dissipation yet included and there are two
conserved quantities:  The reduced energy and the reduced angular momentum,
\begin{eqnarray}
E & = & \frac{1}{2}v^2-\frac{m}{r}+\frac{3}{8}(1-3\eta)v^4+\frac{1}{2
}(3+\eta)v^2\frac{m}{r}+\frac{1}{2}\eta\frac{m}{r}\dot{r}^2+\frac{1}{2}{\left( \frac{m}{r} \right)}^2 \nonumber \\
& + & \frac{5}{16}(1-7\eta+13\eta^2)v^6+\frac{1}{8}(21-23\eta-27\eta^2)\frac{m}{r}v^4 +\frac{1}{4}\eta(1-15\eta)\frac{m}{r}v^2\dot{r}^2 \nonumber \\
& - & \frac{3}{8}\eta(1-3\eta)\frac{m}{r}\dot{r}^4+\frac{1}{8}(14-55\eta+4\eta^2){\left(\frac{m}{r}\right)}^2v^2+\frac{1}{8}(4+69\eta+12\eta^2){\left( \frac{m}{r} \right)}^2\dot{r}^2-\nonumber\\
& - &\frac{1}{4}(2+15\eta){\left( \frac{m}{r} \right)}^3 \label{ee}\\
J& = & ({\bf {r} \times {v}})\left\{ 1+\frac{1}{2}v^2(1-3\eta)+(3+\eta)\frac{m}{r}+\frac{3}{8}(1-7\eta+13\eta^2)v^4\right. \nonumber \\
& + &
\left.\frac{1}{2}(7-10\eta-9\eta^2)\frac{m}{r}v^2-\frac{1}{2}\eta(2+5\eta)\frac{m}{r}\dot{r}^2+\frac{1}{4}(14-41\eta+4\eta^2){\left(
\frac{m}{r}^2\right)}\right\}\label{jj}
\end{eqnarray}

The energy and angular momentum are both measured in units of the 
reduced rest
energy
$\mu=m_2/m$.

For circular orbits $\ddot r_o=\dot r_o=0$ which, from the equations of
motion, requires
	\be
	\dot \phi_o^2={mA_o\over r_o^3}.
	\label{cond2}
	\ee
This condition reduces to a quadratic in $\dot \phi_o^2$ and
we obtain the angular frequency as a
function of the radius of the circular orbit for $\eta>0$:
	\be
	\dot \phi_o^2[ r_o,\eta]={ F-\sqrt{F^2-H}\over K}
	\label{phi2}
	\ee
with
	\ba
	F&=&{1\over 2}\eta \left (13-4\eta\right )\left ({m\over r_o}\right
	)^2+1-\left (1+3\eta\right ){m\over r_o}
        \nonumber \\
	H&=& 4\eta  \left( 3 - 4 \eta  \right) \left ({m\over r_o}\right )^2
            \left(1-2(2+\eta){m\over r_o}+{3\over 4}(12+29\eta)\left
	({m\over r_o}\right )^2
	\right)  
               \nonumber \\
	K&=&{ r_o^2 2\eta   \left( 3 - 4\eta  \right){m\over r_o} }\ \ .\nonumber
	\ea
We show $\dot\phi^2_o(r_o)$ as a function of the circular radius in fig.\
\ref{phidot}. This gives the complete initial
conditions for any circular orbit, stable or unstable.
\begin{figure}
\centerline{\psfig{file=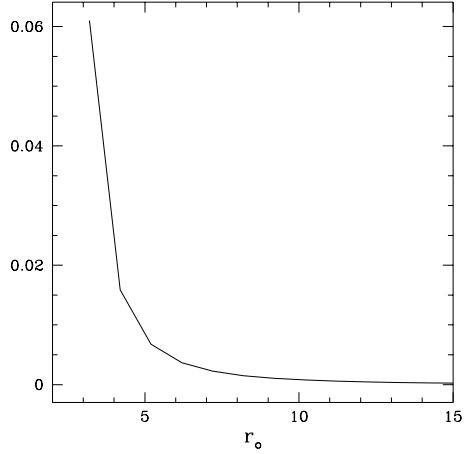,width=2.5in}}
\caption{$\dot\phi^2_o(r_o)$ as a function of the radius of the
circular orbit to 2PN order with $\eta=1/4$.
\label{phidot}}  \end{figure}

To include the effects of dissipation, 
the 5/2PN correction needs to be 
added to $A$ and $B$.  They are respectively
	\ba
	A_\frac{5}{2} & = & -\frac{8}{5}\eta\frac{m}{r}\dot{r}\left[3{v}^2
	+\frac{17}{3}\frac{m}{r}  \right] \\
	B_\frac{5}{2} & = & \frac{8}{5}\eta\frac{m}{r}
	\left[v^2+3\frac{m}{r}\right] .
	\label{52}
	\ea

The instantaneous rate of energy loss to gravitational waves in the
far-zone is calculated in Ref.\ \cite{iyer} for general orbits to 2PN
order.  
We cite their results here.  Defining $L_{GW}= \biggl(\frac{d{
E}}{dt}\biggr )_{\rm far-zone}^{\rm inst} $ in units of $\mu^2$, they derived
\begin{eqnarray}
L_{GW}&=& 
\dot{{ E}}_N + 
\dot{{ E}}_{1PN} + \dot{{ E}}_{2PN}\,,\\
\dot{{ E}}_N &=& {8 \over 15}\frac{m^2}{r^4}
           \left\{ 12v^2 - 11\dot{r}^2\right\} \,,\\
\dot{{ E}}_{1PN} &=&{8 \over 15}\frac{m^2}{r^4}
               \left\{\frac{1}{28}
               \left[(785 - 852\eta)v^4\right.\right.\nonumber \\
              &-& \left.\left.2(1487 - 1392\eta)
	      v^2\dot{r}^2\right.\right.\nonumber \\
              &-& \left.\left.160(17
-\eta) \frac{m}{r}\,v^2\right.\right.\nonumber \\
              &+& \left.\left.3(687 - 620\eta)\dot{r}^4
+ 8 (367 - 15\eta)\frac{m}{r}\,\dot{r}^2\right.\right.\nonumber \\
              &+& \left.\left.16(1 - 4\eta)\,\frac{m^2}{r^2}
\right]\right\}\,,\\
\dot{{ E}}_{2PN} &=&{8 \over 15}\frac{m^2}{r^4}
               \left\{
                  \frac{1}{42} (1692 - 5497\eta
                 + 4430\eta^2)v^6\right.\nonumber \\
              &-& \left.\frac{1}{14} (1719 - 10278\eta
+ 6292\eta^2) v^4\dot{r}^2\right.\nonumber \\
              &-& \left.\frac{1}{21} (4446 - 5237\eta
+ 1393\eta^2)\frac{m}{r}\,v^4\right.\nonumber \\
              &+& \left.\frac{1}{14} (2018 - 15207\eta
+ 7572\eta^2)v^2\dot{r}^4\right.\nonumber \\
              &+& \left.\frac{1}{7} (4987 - 8513\eta
+ 2165\eta^2)\frac{m}{r}\,v^2\dot{r}^2\right.\nonumber \\
              &+& \left.\frac{1}{756} (281473 + 81828\eta
+ 4368\eta^2)\frac{m^2}{r^2}\,v^2\right.\nonumber \\
              &-& \left.\frac{1}{42} (2501 - 20234\eta
+ 8404\eta^2)\dot{r}^6 \right.\nonumber \\
              &-& \left.\frac{1}{63} (33510 - 60971\eta
+ 14290\eta^2)\frac{m}{r}\,\dot{r}^4\right.\nonumber \\
              &-& \left.\frac{1}{252} (106319 + 9798\eta
+ 5376\eta^2) \frac{m^2}{r^2}\,\dot{r}^2\right.\nonumber \\
              &+& \left.\frac{2}{63} (-253 + 1026\eta
- 56\eta^2)\frac{m^3}{r^3}\right\}
\end{eqnarray}
The subscript denotes the order of the term in the PN expansion.

The transverse traceless waveform to $2PN$ order was also computed in
Ref. \cite{iyer} and we write the instantaneous contribution 
here for completeness.  The Earth is
a distance $D$ from the pair and in a direction $\hat N$.  
The vector $\vec r$ is the vector marking the separation of the pair and
$\hat n$ the corresponding unit vector.  The mass difference is
$\delta m=m_1-m_2$.
The notation used is
$n_{ij}=n_in_j$ and, as usual, $n_{(i}v_{j)}=n_{i}v_j+n_jv_i$.
The waveform is then
\begin{eqnarray}
{ {h_{km}}^{TT}}
&=&
\frac{2}{D}P_{ijkm}\left\{ {\xi_{ij}}^{(0)}
+\frac{\delta m}{m}{\xi_{ij}}^{(0.5)} \right. \nonumber \\ 
&+& \left.{\xi_{ij}}^{(1)}+\frac{\delta m}{m}{\xi_{ij}}^{(1.5)}+
{\xi_{ij}}^{(2)}  \right\}\label{htt}
\end{eqnarray}
with the projection defined as
	\begin{equation}
	P_{ijkm}(\hat N)=\left (\delta_{ik}-N_iN_k\right )
	\left(\delta_{jm}-N_jN_m\right )-{1\over 2}
	\left (\delta_{ij}-N_iN_j\right )
	\left(\delta_{km}-N_kN_m\right )
	\end{equation}
and
\begin{eqnarray*}
{\xi_{ij}}^{(0)} &=& 2\left( v_{ij} -\frac{m}{r}n_{ij}  \right) \\
{\xi_{ij}}^{(0.5)} &=&  \left\{ 3({\bf N.n}) \frac{m}{r} \left[2n_{(i}v_{j)}-\dot{r}n_{ij} \right] + ({\bf N.v}) \left[ \frac{m}{r}n_{ij} -2v_{ij} \right] \right\}\\
{\xi_{ij}}^{(1)} &=& \frac{1}{3} \left\{ (1-3\eta)\left[ ({\bf N.n})^2 \frac{m}{r} \left( \left(3v^2-15\dot{r}^2 +7\frac{m}{r}\right)n_{ij}+30\dot{r}n_{(i}v_{j)} -14v_{ij} \right) \right. \right. \\ 
&+&\left. \left. ({\bf N.n})({\bf N.v})\frac{m}{r}\left[ 12\dot{r}n_{ij}-32n_{(i}v_{j)} \right]+({\bf N.v})^2 \left[6v_{ij}-2\frac{m}{r}n_{ij} \right] \right]    \right.\\
&+& \left. \left[3(1-3\eta)v^2 - 2(2-3 \eta)\frac{m}{r}\right] v_{ij} +4\frac{m}{r} \dot{r}(5+3\eta)n_{(i}v_{j)} \right.\\
&+&
\left.
\frac{m}{r}\left[3(1-3\eta)\dot{r}^2-(10+3\eta)v^2+29\frac{m}{r}
\right]n_{ij}\right\}\\
{\xi_{ij}}^{(1.5)}&=& \frac{1}{12}(1-2 \eta) \left\{({\bf N.n})^3 \frac{m}{r}\left[(45v^2-105\dot{r}^2+90\frac{m}{r})\dot{r}n_{ij} -96\dot{r}v_{ij} \right.\right.\\
&-& \left. \left.\left(42v^2-210\dot{r}^2+88\frac{m}{r}\right) n_{(i}v_{j)} \right] \right.\\
&-&\left. ({\bf N.n})^2({\bf N.v})\frac{m}{r}\left[\left(27v^2-135\dot{r}^2+84\frac{m}{r}\right)n_{ij}+336\dot{r}n_{(i}v_{j)}-172v_{ij} \right]    \right.\\
&-& \left.({\bf N.n})({\bf N.v})^2 \frac{m}{r}\left[48\dot{r}n_{ij}-184n_{(i}v_{j)}  \right]+({\bf N.v})^3\left[4\frac{m}{r}n_{ij}-24v_{ij} \right] \right\}\\
&-& \frac{1}{12}({\bf N.n}) \frac{m}{r} \left\{ \left[ (69-66 \eta)v^2-(15-90\eta)\dot{r}^2-(242-24\eta)\frac{m}{r}\right] \dot{r}n_{ij} \right.\\
&-& \left.\left[(66-36\eta)v^2+(138+84\eta)\dot{r}^2 \right. \right. \\
&-& \left. \left. (256-72\eta)\frac{m}{r}\right] n_{(i}v_{j)}+(192+12\eta)\dot{r}v_{ij} \right\}\\
&+& \frac{1}{12} ({\bf N.v}) \left\{\left[(23-10\eta)v^2-(9-18\eta)\dot{r}^2-(104-12\eta)\frac{m}{r}\right] \frac{m}{r}n_{ij} \right. \\
&-& \left. (88+40\eta)\frac{m}{r}\dot{r}
n_{(i}v_{j)} -
\left[(12-60\eta)v^2-(20-52\eta)\frac{m}{r}\right] v_{ij}\right\} \\
{\xi_{ij}}^{(2)} &=& \frac{1}{120}(1-5\eta+5\eta^2) \left\{240({\bf N.v})^4v_{ij}-({\bf N.n})^4\frac{m}{r}  \right. \\
&\times & \left. \left[\left(90v^4 +(318\frac{m}{r}-1260\dot{r}^2)v^2+344 \frac{m^2}{r^2} +1890\dot{r}^4 \right. \right. \right. 
- \left. \left. \left. 2310 \frac{m}{r}\dot{r}^2 \right) n_{ij}   \right.  \right. \\
&+& \left. \left. \left(1620v^2 +3000\frac{m}{r} -3780\dot{r}^2  \right)\dot{r}n_{(i}v_{j)} -\left(336v^2-1680\dot{r}^2+668\frac{m}{r}\right)v_{ij} \right] \right. \\
&-& \left. ({\bf N.n})^3({\bf N.v})\frac{m}{r}\left[\left(1440v^2-3360\dot{r}^2+3600\frac{m}{r}\right)\dot{r}n_{ij} \right. \right. \\
&-& \left. \left.  \left(1608v^2-8040\dot{r}^2+3864\frac{m}{r}\right)n_{(i}v_{j)} -3960\dot{r}v_{ij} \right]  \right. \\
&+& \left. 120({\bf N.v})^3({\bf N.n})\frac{m}{r}\left( 3\dot{r}n_{ij}-20 n_{(i}v_{j)}\right) \right. \\
&+& \left. ({\bf N.n})^2 ({\bf N.v})^2 \frac{m}{r} \left[\left(396v^2-1980\dot{r}^2+1668\frac{m}{r}\right)n_{ij}+6480\dot{r}n_{(i}v_{j)} \right. \right. \\
&-&  \left. \left. 3600v_{ij}  \right] \right\} -\frac{1}{30}({\bf N.v})^2  \left\{ \left[(87-315\eta+145\eta^2)v^2-(135-465\eta+75\eta^2)\dot{r}^2 \right. \right. \\
&-& \left. \left. (289-905\eta+115\eta^2)\frac{m}{r}  \right]\frac{m}{r}n_{ij} \right. \\
&-& \left.\left(240-660\eta-240\eta^2\right) \dot{r} n_{(i}v_{j)} \right. \\
&-&  \left. \left[ (30-270\eta +630\eta^2)v^2-60(1-6\eta+10\eta^2)\frac{m}{r}  \right]v_{ij}   \right\}  \\
&+& \frac{1}{30} ({\bf N.n})({\bf N.v}) \frac{m}{r} \left\{ \left[ (270-1140\eta +1170\eta^2)v^2 \right. \right. \\
&-& \left. \left. (60-450\eta +900\eta^2)\dot{r}^2 -(1270-3920\eta+360\eta^2)\frac{m}{r}\right]\dot{r}n_{ij} \right. \\ 
&-& \left.\left[(186-810\eta+1450\eta^2 )v^2+(990-2910\eta-930\eta^2)\dot{r}^2  \right. \right. \\
&-&  \left. \left. (1242-4170\eta+1930\eta^2)\frac{m}{r} \right] n_{(i}v_{j)} \right. \\
&+&  \left. \left[1230-3180\eta -90\eta^2  \right] \dot{r}v_{ij} \right\} \\
&+& \frac{1}{60} ({\bf N.n})^2 \frac{m}{r} \left\{ \left[ (117-480\eta+540\eta^2)v^4-(630-2850\eta+4050\eta^2)v^2\dot{r}^2 \right. \right. \\
&-& \left. \left. (125-740\eta +900\eta^2)\frac{m}{r}v^2 \right. \right. \\
&+&  \left. \left. (105-1050\eta+3150\eta^2)\dot{r}^4+(2715-8580\eta+1260\eta^2)\frac{m}{r}\dot{r}^2 \right. \right. \\
&-& \left. \left. (1048-3120\eta+240\eta^2)\frac{m^2}{r^2} \right]n_{ij} \right. \\
&+& \left. \left[(216-1380\eta+4320\eta^2)v^2+(1260-3300\eta-3600\eta^2)\dot{r}^2 \right. \right. \\
&-& \left. \left. (3952-12860\eta+3660\eta^2)\frac{m}{r}  \right] \dot{r}n_{(i}v_{j)}   \right. \\
&-& \left. \left[(12-180\eta+1160\eta^2)v^2+(1260-3840\eta-780\eta^2)\dot{r}^2 \right. \right. \\
&-& \left. \left. (664-2360\eta+1700\eta^2)\frac{m}{r}\right]v_{ij} \right\} \\
&-& \frac{1}{60} \left\{ \left[(66-15\eta-125\eta^2)v^4 \right. \right. \\
&+& \left. \left. (90-180\eta-480\eta^2)v^2\dot{r}^2 -(389+1030\eta-110\eta^2)\frac{m}{r}v^2 \right. \right. \\
&+& \left. \left.(45-225\eta+225\eta^2)\dot{r}^4+(915-1440\eta+720\eta^2)\frac{m}{r}\dot{r}^2 \right. \right. \\
&+& \left. \left. (1284+1090\eta)\frac{m^2}{r^2}\right]\frac{m}{r}n_{ij} \right. \\
&-& \left. \left[(132+540\eta-580\eta^2)v^2+(300-1140\eta+300\eta^2)\dot{r}^2 \right. \right. \\
&+& \left. \left.(856+400\eta +700\eta^2)\frac{m}{r}\right]\frac{m}{r}\dot{r}n_{(i}v_{j)}\right. \\
&-&  \left. \left[(45-315\eta+585\eta^2)v^4+(354-210\eta-550\eta^2)\frac{m}{r}v^2 \right. \right. \\
&-& \left. \left. (270-30\eta+270\eta^2)\frac{m}{r}\dot{r}^2 \right. \right. \\
&-& \left. \left. (638+1400\eta-130\eta^2)\frac{m^2}{r^2} \right]v_{ij} \right\}
\end{eqnarray*}
The superscript denotes the order in the PN expansion.  A full
explanation of the derivation of these terms can be found in the
original Reference \cite{iyer}.

}

\end{document}